\documentclass[prb,aps,amsmath,amssymb,showpacs,twocolumn]{revtex4-1}
\usepackage{amssymb,graphicx,float,dcolumn,bm,subfigure,color}
\usepackage[titletoc,title]{appendix}
\usepackage[dotinlabels]{titletoc}
\usepackage{subfigure}
\usepackage{braket}
\usepackage{nicefrac}

\newcommand{\be}{\begin{equation}}
\newcommand{\ee}{\end{equation}}

\newcommand{\ba}{\begin{eqnarray}}
\newcommand{\ea}{\end{eqnarray}}

\newcommand{\E}{{\cal E}}

\renewcommand{\vec}[1]{{\textbf{\textit{#1}}}}

\begin{document}

\title{Density functional theory of the fractional quantum Hall effect}

\author{Jianyun Zhao$^1$, Manisha Thakurathi$^2$, Manish Jain$^3$, Diptiman Sen$^2$, J. K. Jain$^1$}
\affiliation{$^1$Physics Department, 104 Davey Laboratory, Pennsylvania State University, University Park, PA 16802}
\affiliation{$^2$Centre for High Energy Physics, Indian Institute of Science, Bengaluru 560012, India}
\affiliation{$^3$Department of Physics, Indian Institute of Science, Bengaluru, 560012, India}

\begin{abstract}
A conceptual difficulty in formulating the density functional theory of the fractional quantum Hall  effect is that while in the standard approach the Kohn-Sham orbitals are either fully occupied or unoccupied, the physics of the fractional quantum Hall effect calls for fractionally occupied Kohn-Sham orbitals. This has necessitated averaging over an ensemble of Slater determinants to obtain meaningful results. We develop an alternative approach in which we express and minimize the grand canonical potential in terms of the composite fermion variables. This provides a natural resolution of the fractional-occupation problem because the fully occupied orbitals of composite fermions automatically correspond to fractionally occupied orbitals of electrons. We demonstrate the quantitative validity of our approach by evaluating the density profile of fractional Hall edge as a function of temperature and the distance from the delta dopant layer and showing that it reproduces edge reconstruction  in the expected parameter region.
\end{abstract}

\maketitle

The density functional theory (DFT) is a powerful tool for treating many particle ground states.  A quantitatively reliable DFT of the fractional quantum Hall (FQH) effect would obviously be extremely useful for elucidating the fundamental physics of FQH systems with spatially varying density, whether induced by an external potential or generated spontaneously, which are not readily amenable to many of the theoretical methods used in the field.  However, the problem is  nontrivial\cite{Ferconi95,Heinonen95} because the solution is not close to a single Slater determinant in which some of the Kohn-Sham orbitals are fully occupied and the others empty, but instead entails fractional occupation of Kohn-Sham orbitals, as demanded by the physics of the FQH effect (FQHE). Theoretically, fractionally occupied orbitals arise because all single particle orbitals of electrons are degenerate in the absence of interaction, and interaction produces a strongly correlated state in a nonperturbative fashion.  A possible way to obtain on-average fractionally filled Kohn-Sham orbitals is through ensemble averaging. In the first application of DFT to the FQHE, Ferconi, Geller and Vignale\cite{Ferconi95} averaged over a thermal ensemble to achieve fractional fillings and obtained the density profile at the edge in the presence of a confinement potential. In another approach, Heinonen, Lubin and Johnson~\cite{Heinonen95} performed an average over the ensemble of Slater determinants obtained in successive steps of the iterative scheme for solving the Kohn-Sham equations, and also generalized their approach to include the spin degree of freedom~\cite{Lubin97,Heinonen99}. 

We present in this work a formulation of the DFT of FQHE in terms of composite fermions rather than electrons. This provides a natural solution to the fractional-occupation problem, because occupied orbitals of composite fermions, as obtained in the DFT formulation, automatically correspond to fractionally filled Kohn-Sham orbitals of electrons. We minimize, in a local density approximation, the thermodynamic potential expressed as a functional of the CF density in various CF Landau levels, using an exchange correlation functional for composite fermions deduced from microscopic calculations and an entropy functional that properly incorporates the physics of strong correlations. To test the quantitative validity of our approach, we determine the density profile of the FQHE edge and find, in agreement with previous exact diagonalization studies, that the edge undergoes a reconstruction when the delta-dopant layer containing the positive neutralizing charge is farther than a critical distance.  We further find that, for general fractions, edge reconstruction extends much deeper into the interior of the sample than previously suspected, and determine the temperatures where it is washed out by thermal fluctuations. As another application, we calculate how the periodic potential produced by a Wigner crystal (WC) in a nearby layer affects the density of composite fermions at $\nu\approx 1/2$. 

The objective is to minimize the grand potential 
\be
\Omega[\rho]=\E_{\rm xc}[\rho]+\E_{\rm H}[\rho]+\int d^2 \vec{r} \rho(\vec{r})(V(\vec{r})-\mu)-k_{\rm B}TS
\label{Vignale1}
\ee
expressed in terms of the electron density $\rho(\vec{r})$, which is related to the local electron filling factor $\nu(\vec{r})$ as $\rho(\vec{r})=\nu(\vec{r})/2\pi l^2$, where $l=\sqrt{\hbar c/eB}$ is the magnetic length. Here $\E_{\rm xc}$ and $\E_{\rm H}$ are the exchange-correlation and Hartree energies, $V(\vec{r})$ is the potential energy due to interaction with an external charge distribution, $\mu$ is the chemical potential, $T $ is the temperature, and $S$ is the entropy. 
To express $\Omega[\rho]$ in terms of composite fermions, let us recall some relevant facts about composite fermions\cite{Jain89,Jain07}. The density of composite fermions is the same as that of electrons, but composite fermions experience an effective magnetic field $B^*=B-2\rho\phi_0$ ($\phi_0=hc/e$), form Landau-like levels [called $\Lambda$ levels ($\Lambda$Ls)], and their filling factor $\nu^*$ is related to the electron filling factor by the equation $\nu=\nu^*/(2\nu^*\pm 1)$. (We specialize, for simplicity, to composite fermions carrying two flux quanta.) Because we will deal with non-uniform densities, we define $\nu^*(\vec{r})=\sum_j \nu_j^*(\vec{r})$, where $\nu_j^*(\vec{r})$ is the local filling factor of the $j$th $\Lambda$L. The effective CF cyclotron energy is given by the relation 
$ 
\hbar \omega_c^*=\hbar {eB^*\over m^*c}=\hbar {eB\over (2\nu^*\pm 1) m^*c}\equiv 
{\alpha\over 2\nu^*\pm 1}\; {e^2 \over \epsilon l}
$
where the last equality is motivated from dimensional arguments\cite{Halperin93,Jain07}, and has also been tested in calculations that identify the CF cyclotron energy to the energy required to excite a far separated CF particle-hole pair\cite{Scarola02}. Explicit calculation yields $\alpha=0.33$ for a system with zero thickness\cite{Halperin93,Jain07}, which is what we shall assume below.

\begin{figure}[t]
\begin{center}			
\hspace{-3mm}
\includegraphics[scale=0.35]{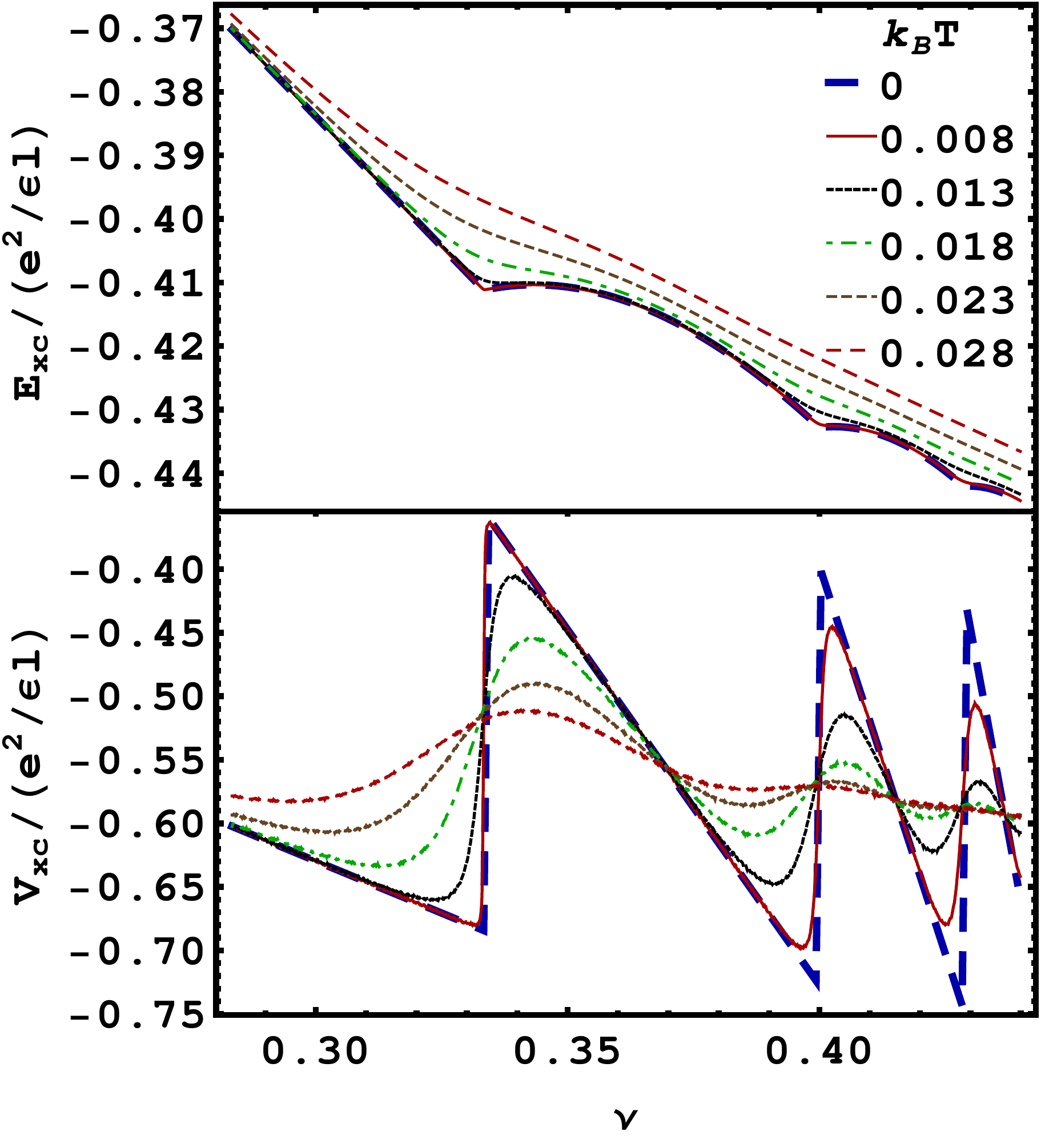}
\end{center}
\vspace{-8mm}
\caption{Exchange-correlation energy $E_{\rm xc}(\nu)$ (Eq.~\ref{Exc}) and potential $V_{\rm xc}(\nu)=E_{\rm xc}+\nu \partial E_{\rm xc} / \partial \nu$ as a function of the filling factor $\nu$ for several temperatures. 
}
\label{Exc_plot}		
\end{figure}

We first determine the exchange correlation function by making the local density approximation, which is valid when the variation in the density is sufficiently slow that we can consider it to be locally constant. In other words, we assume that the variations in density are negligible on the scale of the CF magnetic length $l^*=\sqrt{\hbar c/e|B^*|}$. We write
\be
\E_{\rm xc}=\int d^2 \vec{r} \rho(\vec{r}) E_{\rm xc}[\rho(\vec{r})]
\ee
where $E_{\rm xc}[\rho(\vec{r})]$ is the exchange-correlation energy per particle for a system with {\em uniform} density.  For a uniform system, $E_{\rm xc}$ is precisely the energy that is usually obtained in numerical calculations (because the total energy includes electron-background and background-background terms which cancel the Hartree part of the interaction energy of electrons). It is possible to obtain, in the CF theory, the thermodynamic limits for the energies at the discrete value of fillings $\nu=n/(2n\pm1)$, where the electronic ground states are accurately represented as $\nu^*=n$ filled $\Lambda$Ls of composite fermions \cite{Jain97,Jain97b}.  From explicit calculation with the microscopic theory of composite fermions, the exchange-correlation energy per {\em electron} at $\nu=n/(2n\pm1)$ is given very accurately by\cite{BalramJain}
\be
E_{\rm xc}\left[\nu={n\over 2n\pm 1}\right]=a {n\over 2n\pm 1} +b 
\label{Ajit}
\ee
with $a=-0.324$ and $b=-0.303$. (We express all energies and also $k_{\rm B}T$ in units of $e^2/\epsilon l$, which is $\sim 150$K at $B=9$T for parameters appropriate for GaAs.) The energy as a function of continuous $\nu$ has downward cusps at $\nu=n/(2n\pm1)$. Rather than attempting a microscopic calculation for the full curve of energy vs. filling factor, which can be performed assuming that the composite fermions in the partially filled $\Lambda$L form a crystal \cite{Archer13}, we will make a model that is more natural from the DFT point of view and sufficient for current purposes. We will interpret the exchange-correlation energy of electrons as a sum of exchange-correlation and kinetic energies for composite fermions:
\be
E_{\rm xc}[\nu]=E^{*}_{\rm xc}[\nu]+E^{*}_{\rm K}[\nu]
\label{ExcElectron}
\ee
where we follow the usual convention that all quantities marked by an asterisk $*$ refer to composite fermions. We shall further assume that composite fermions themselves are weakly correlated, i.e., $E^{*}_{\rm xc}$ is smooth and all cusps arise from $E^{*}_{\rm K}$. 
In terms of the CF cyclotron energy $\hbar \omega_c^*$, $E^{*}_{\rm K}$ at the special fillings $\nu=n/(2n\pm 1)$ is given by
$E^{*}_{\rm K}\left[\nu={n\over 2n\pm 1}\right]={n\over 2}\;\hbar \omega_c^*={\alpha \over 2}\;{n\over 2n\pm 1}$.
This leads us to the final form of $E_{\rm xc}$ for arbitrary $\nu$ that we use in our calculations below:
\be
E_{\rm xc}[\nu]=a \nu + b -{\alpha \over 2}\;\nu+E^{*}_{\rm K}[\nu]
\label{Exc}
\ee
At T$=0$, the average kinetic energy per CF for a general filling $\nu=\nu^*/(2\nu^*\pm 1)$ with $n= {\rm int} (\nu^*) $ is given by 
\be 
E^{*}_{\rm K}[\nu]= \left(2n+1-{n(n+1)\over\nu^{*}}\right)|1-2\nu|{\alpha\over2}
\label{kinetic1}
\ee  
where $\nu^*=|{\nu\over1-2\nu}|$. At finite T, the CF kinetic energy per particle can be evaluated numerically as $E^*_{\rm K}={ 1\over \nu^*} \sum_j (j+1/2)\hbar \omega_c^* \nu_j^*$ with $\nu^*=\sum_j\nu_j^*$ and $\nu_j^* = (e^{{(j+1/2)\hbar\omega_c^* -\mu\over k_{\rm B}T}}+1)^{-1}$. 
The resulting $E_{\rm xc}$ is plotted in Fig.~\ref{Exc_plot} along with $V_{\rm xc}=\delta \E_{\rm xc}/\delta \nu(\vec{r})=E_{\rm xc}+\nu \partial E_{\rm xc} / \partial \nu$. In the limit of $T=0$, $E_{\rm xc}$ has cusps and $V_{\rm xc}$ discontinuities at $\nu=n/(2n\pm 1)$. We note that, for simplicity, we have not incorporated into our model the physics of the $\nu=n/(4n\pm 1)$ incompressible states at $\nu<1/3$ described in terms of composite fermions carrying four flux quanta. 

To obtain an expression for the entropy, we need the knowledge of the excitation spectrum of the strongly correlated FQH state. As detailed calculations have shown \cite{Balram13}, the counting of excited states is consistent with the model of weakly interacting composite fermions for temperatures small compared to the CF Fermi energy $E^*_{\rm F}\sim 0.1 e^2/\epsilon l$.  We note that the degeneracy of the CF $\Lambda$Ls is determined by the effective magnetic field, which in turn depends on density and thus position. As a result, a sum over all single CF energy levels is written as 
\be
\sum_{i, \gamma}=\sum_{i} \int {d^2 \vec{r} \over 2\pi (l^*)^2} =  \sum_{i}{1\over2\pi l^{2}}\int d^2 \vec{r}  |1-2\nu(\vec r)|
\ee
where $i$ is the $\Lambda$L index and $\gamma$ labels single CF states within a $\Lambda$L. The entropy of composite fermions is thus given by
\be
\begin{split}
S[\{\nu_{i}^{*}\}]=-{1\over2\pi l^{2}}\int d^2 \vec{r}  |1-2\nu(\vec r)|\\
\sum_{i}\big\{\nu_{i}^{*}(\vec r)\ln[\nu_{i}^{*}(\vec r)]+(1-\nu_{i}^{*}(\vec r))\ln[1-\nu_{i}^{*}(\vec r)]\big\}
\end{split}
\label{entropy}
\ee
For FQH states corresponding to filled $\Lambda$Ls ($\nu_i=1$ or $0$) the entropy vanishes as it should. 

 \begin{figure}[t]
\hspace{-2.4mm}
\resizebox{0.164\textwidth}{!}
{\includegraphics{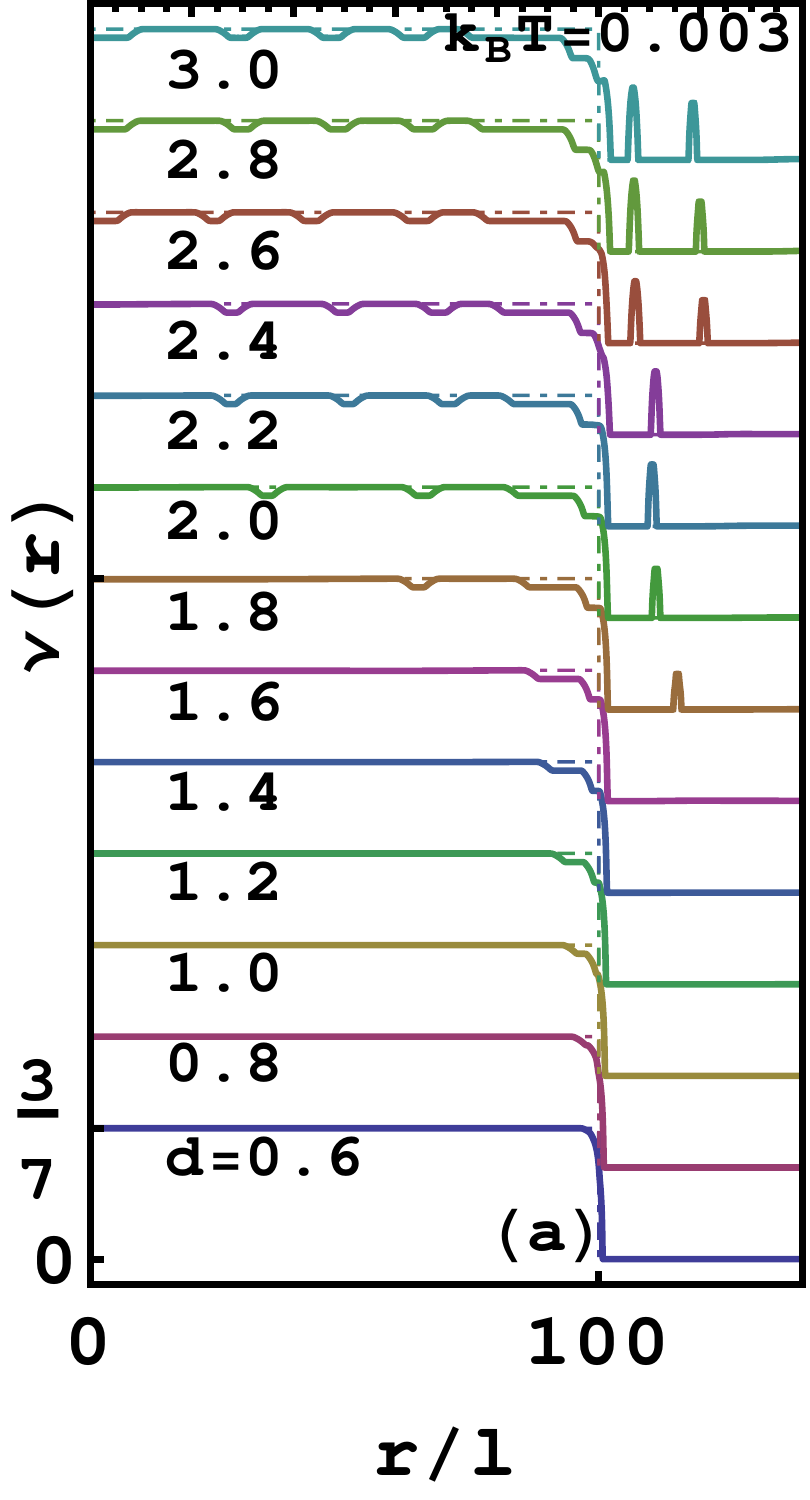}}
\hspace{-2.4mm}
\resizebox{0.164\textwidth}{!}
{\includegraphics{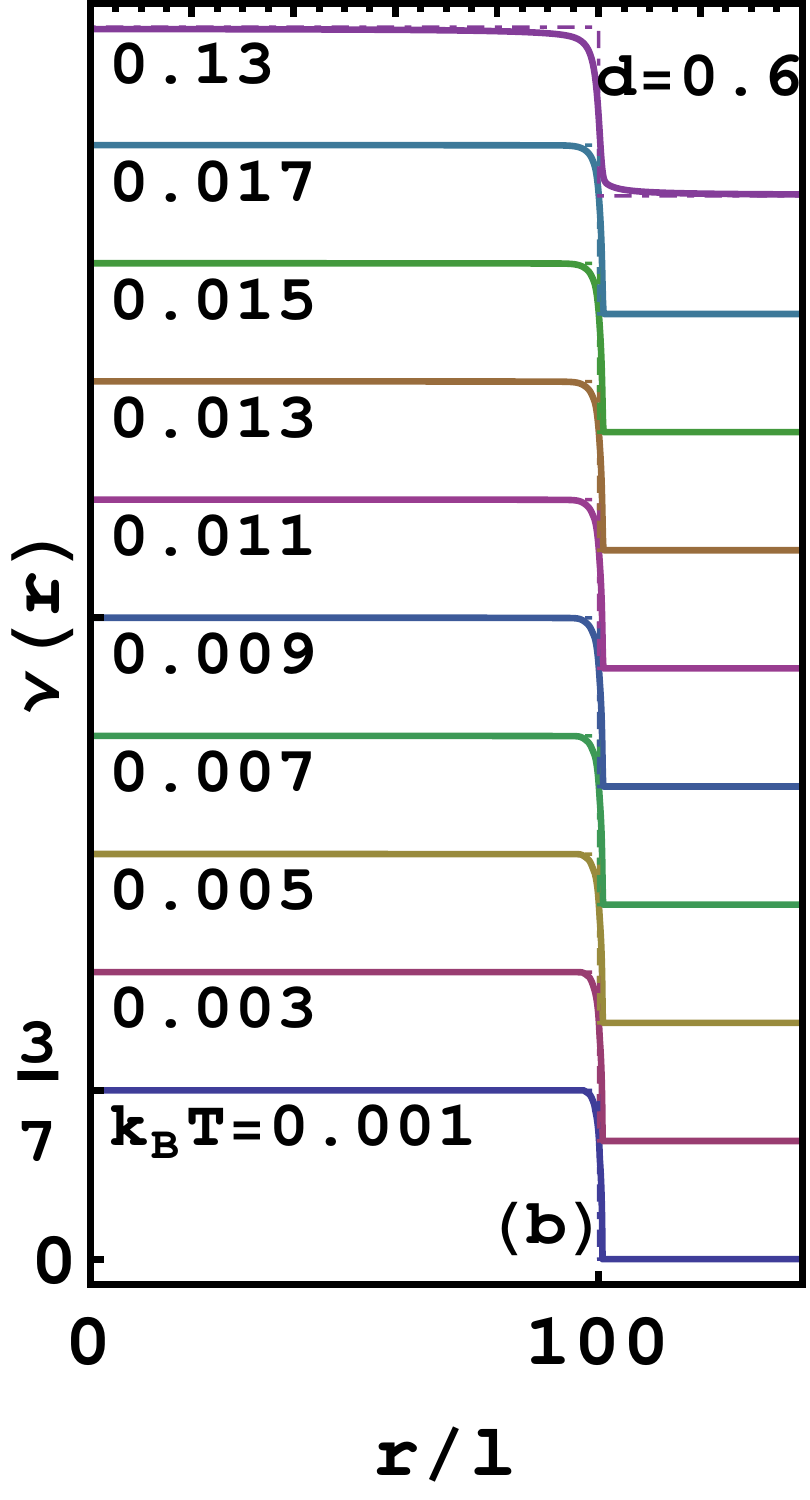}}
\hspace{-2.4mm}
\resizebox{0.164\textwidth}{!}
{\includegraphics{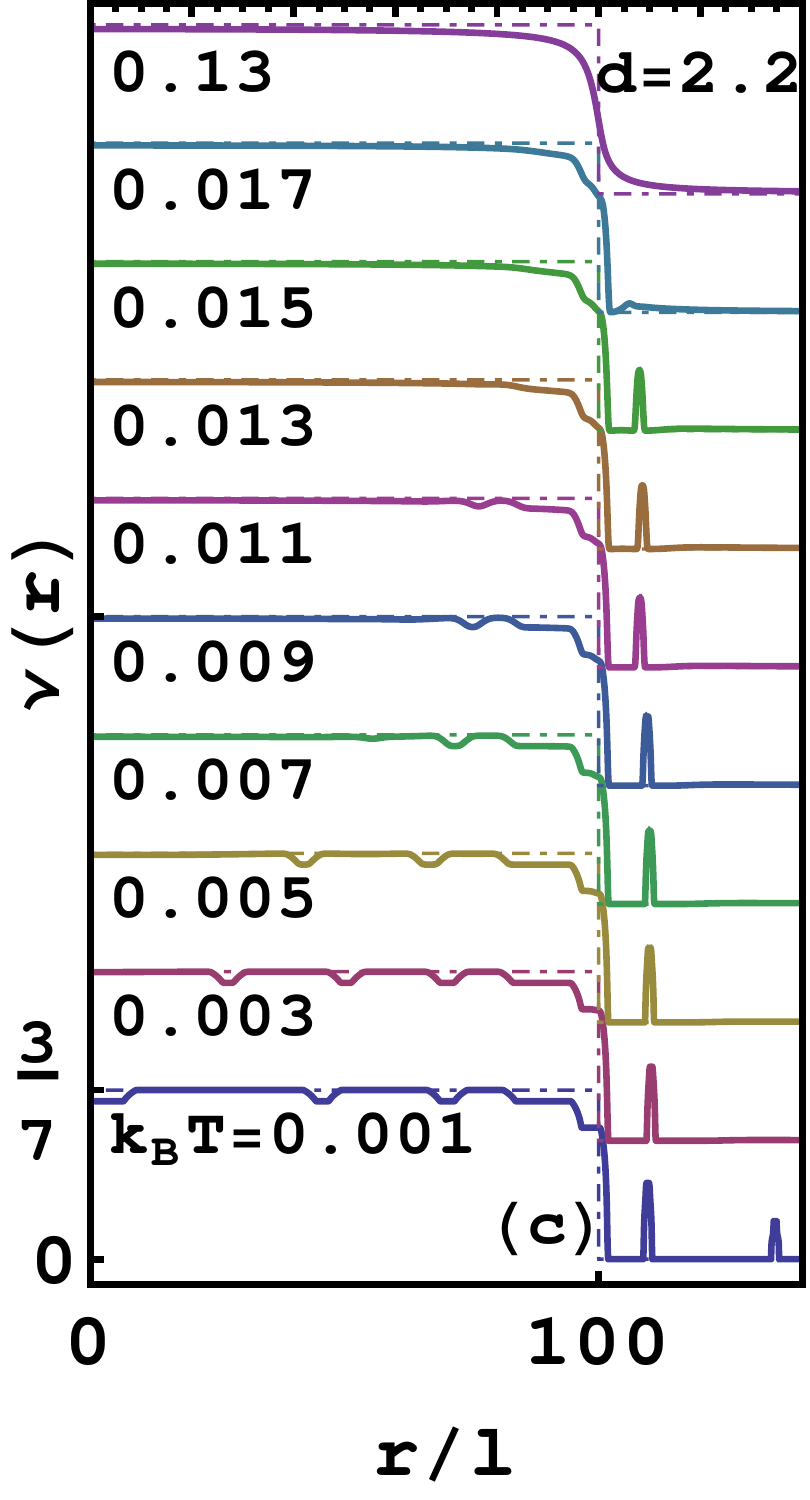}}
\vspace{-7.3mm}
\caption{(a) Evolution of the 3/7 edge as a function of the setback distance $d$ for a small temperature $k_{\rm B}T=0.003$.  The $\nu(\vec{r})$ for successive $d$ are vertically displaced for clarity. Edge reconstruction is seen to occur at $d\approx 1.7$. (b-c) Evolution of the 3/7 edge as a function of temperature for two values of $d$. For $d=2.2$ the edge structure melts at $k_{\rm B}T\approx 0.017$. $k_{\rm B}T$ is quoted in units of $e^2/\epsilon l$.} 
\label{37edge}		
\end{figure}

In terms of $\nu^*_j(\vec{r})$, the thermodynamic potential is rewritten as (with $\E$ representing the total energies and $E$ representing the energies per particle)
\begin{eqnarray}
\Omega[\nu^*_j(\vec r)]&=&\E_{\rm K}^{*}+\E_{\rm H}+\E_{\rm xc}^{*}+ \int {d^2 \vec{r}\over 2\pi l^2}\nu(\vec r)V(\vec r)\nonumber \\
& - &\mu\left(\int {d^2 \vec{r}\over 2\pi l^2}\nu(\vec r)- N\right)-k_{\rm B}TS[\{\nu_{j}^{*}(\vec r)\}]
\label{new_omega}
\end{eqnarray}
where under local density approximation, we have
\be
\E_{\rm K}^{*}[\{\nu_{j}^{*}\}]={\alpha\over2\pi l^{2}} \int d^2 \vec{r} |1-2\nu(\vec r)|^2\sum_{i}\nu_{j}^{*}(\vec r)
\left(j+{1\over2}\right),
\label{kinetic2}
\ee
\be
\E_{\rm xc}^{*}[\nu(\vec r)]={1\over2\pi l^2}\int d^2 \vec{r}\left[\left(a-{\alpha\over2}\right)\nu(\vec r)+b\right]\nu(\vec r),
\ee
and the entropy is given in Eq.~\ref{entropy}. Eq.~\ref{kinetic2} reduces to Eq.~\ref{kinetic1} in the limit of zero temperature, when all $\Lambda$Ls other than the topmost one are fully occupied, but we allow occupation of higher $\Lambda$Ls, as appropriate at finite temperatures. 
The term $\mu N$, where $N$ is the number of electrons, has no effect on the self-consistency equations. The electron density (or filling factor) is given by 
\be
\rho(\vec{r})={\nu(\vec{r})\over 2\pi l^2}={1\over 2\pi l^2} \;{\sum_j \nu_j^*(\vec{r})\over 2 \sum_j \nu_j^*(\vec{r})\pm 1}
\ee
 
Using  
${{\delta\nu(\vec r)}/{\delta\nu_{i}^{*}(\vec r')}}=(1-2\nu(\vec r))|1-2\nu(\vec r)| \delta(\vec{r}-\vec{r}')$
we minimize Eq.~\ref{new_omega} with respect to $\nu_{j}^{*}(\vec r)$. This results in the condition
\be
\nu_j^*(\vec r)={1\over\exp[\epsilon^*_j(\vec{r})/k_{\rm B}T]+1}
\label{central_lambda}
\ee
where $\epsilon^*_j(\vec{r})$, the local self-consistent energy of the $j$th $\Lambda$L, is given by
\be
\epsilon^*_j=|1-2\nu(\vec r)| (\varepsilon_{1j}^{*} + 2k_{\rm B}T s^*) + (1-2\nu(\vec r))(\varepsilon_{2}^{*}-\mu)
\ee
\be
\varepsilon_{1j}^{*}=\left(j+{1\over2}\right)\alpha-4\alpha|1-2\nu(\vec r)|\sum_{i}\left(i+{1\over2}\right)\nu_i^*(\vec r)
\ee
\be 
\varepsilon_{2}^{*}={1\over2\pi l^2}\int{\nu(\vec r')\over|\vec r-\vec r'|}d^2 \vec{r}'+V(\vec r)+2\left(a-{\alpha\over2}\right)\nu(\vec r)+b
\ee
\be
s^{*}=-\sum_{i}\{\nu_{i}^{*}(\vec r)\ln[\nu_{i}^{*}(\vec r)]+(1-\nu_{i}^{*}(\vec r))\ln[1-\nu_{i}^{*}(\vec r)]\}
\ee
The solution $\nu_j^*(\vec{r})$ is obtained by demanding self-consistency of Eq.~\ref{central_lambda}.  From the knowledge of $\nu^*_j(\vec{r})$, the electron density and the free energy $\Omega$ can be readily evaluated.  The self-consistent $\Lambda$L energies $\epsilon^*_j(\vec{r})$ are very complicated functions of various parameters, and display a non-trivial dependence on the position. 

To obtain the self-consistent solution we begin with an initial choice for $\nu^*_j(\vec{r})$ that tracks the neutralizing charge and calculate the new values according to Eq.~\ref{central_lambda} fixing the chemical potential to ensure the correct total charge. A new choice is then obtained by mixing the input and output values, and the procedure is iterated until self-consistency is achieved.  See Supplemental Material (SM)\cite{DFT-SM} for further details. To ensure smoothness on the scale of $l^*$, which is expected on physical grounds and also assumed in local density approximation, we average the local filling factor $\nu^*_j(\vec{r})$  over a length $l_{\rm ave}$ at each step of our self-consistency loop. In our calculations shown below, we use $l_{\rm ave}=l^*$ (which depends on the local filling factor). 
As mentioned above, in the limit $T\rightarrow 0$ the local CF filling factor $\nu^*_j(\vec{r})$  approaches either 0 or 1 in each $\Lambda$L, depending on whether the self-consistent $\Lambda$L energy $\epsilon^*_j(\vec{r})$ is positive or negative. This  produces a fractional value for the local $\nu(\vec{r})$, as appropriate for the physics of the problem. 

As a first application of the above formalism, we consider the behavior at the edge of a FQH state. Following the typical experimental geometry, we shall model the positively charged background as a uniformly charged $\delta$-doped disk at a set-back distance $d$ from the plane containing the electrons.  This corresponds to 
\be
V(\vec{r})=-e^2\int d^2 \vec{r}' {\rho_b(\vec{r}') \over \sqrt{|\vec{r}-\vec{r}'|^2+d^2}}
\label{Vr}
\ee
Exact diagonalization studies on small systems\cite{Wan02, Wan03,Jolad09,Zhang14b} at $\nu=1/3$ have found that an edge excitation mode becomes soft when $d$ becomes larger than a critical value $\sim 1.5 l$ (recall $l\approx 8$nm for $B=9$T), which is interpreted in terms of an edge reconstruction\cite{Chamon94}. The systems were too small to shed light on the nature of the reconstructed edge, or to study this physics at more general fillings of the type $\nu=n/(2n\pm 1)$ which are expected to have much more complex edges. As seen in SM\cite{DFT-SM}, our DFT method shows that for the 1/3 state edge reconstruction occurs at $d=1.5$ at small T, which is a strong confirmation of the quantitative validity of our approach. We illustrate the power of our approach by taking the example of the edge of 3/7 FQH state.  Fig.~\ref{37edge}(a) displays the evolution of the edge at a low temperature as a function of $d$. Edge reconstruction is seen at $d\sim 1.7 l$. Incompressible stripes of $\nu=2/5$ and $\nu=1/3$ are seen to emerge except for very small $d$, with an stripe pattern alternating between 3/7 and 2/5 extending deep into the interior at large $d$ at low T. 
We note that the alternating stripe pattern is qualitatively distinct from that seen in integer quantum Hall effect\cite{Chklovskii92}. The reason is because the densities for nearby FQH states are very close, and thus the stripe formation does not entail a high Hartree cost. Figs.~\ref{37edge}(b) and (c) and S2\cite{DFT-SM} display the evolution of the 3/7 edge as a function of T. Edge reconstruction is absent at small $d$, while for $d=2.2$, it is washed out by $k_{\rm B}T=0.017$, which is much smaller than the CF Fermi energy $E^*_{\rm F}\sim$0.1. Fig.~S3 depicts the spatial dependence of $\epsilon^*_j(\vec{r})$ for several choices of parameters. We note that significant experimental progress has been made toward imaging the quantum Hall edges to explore compressible and incompressible stripes as well as edge reconstruction (see Refs.~\cite{Weis11,Paradiso12,Pascher14,Sabo17} and references therein). 

\begin{figure}[t]
\begin{center}			
\hspace{-2.9mm}
\resizebox{0.164\textwidth}{!}
{\includegraphics{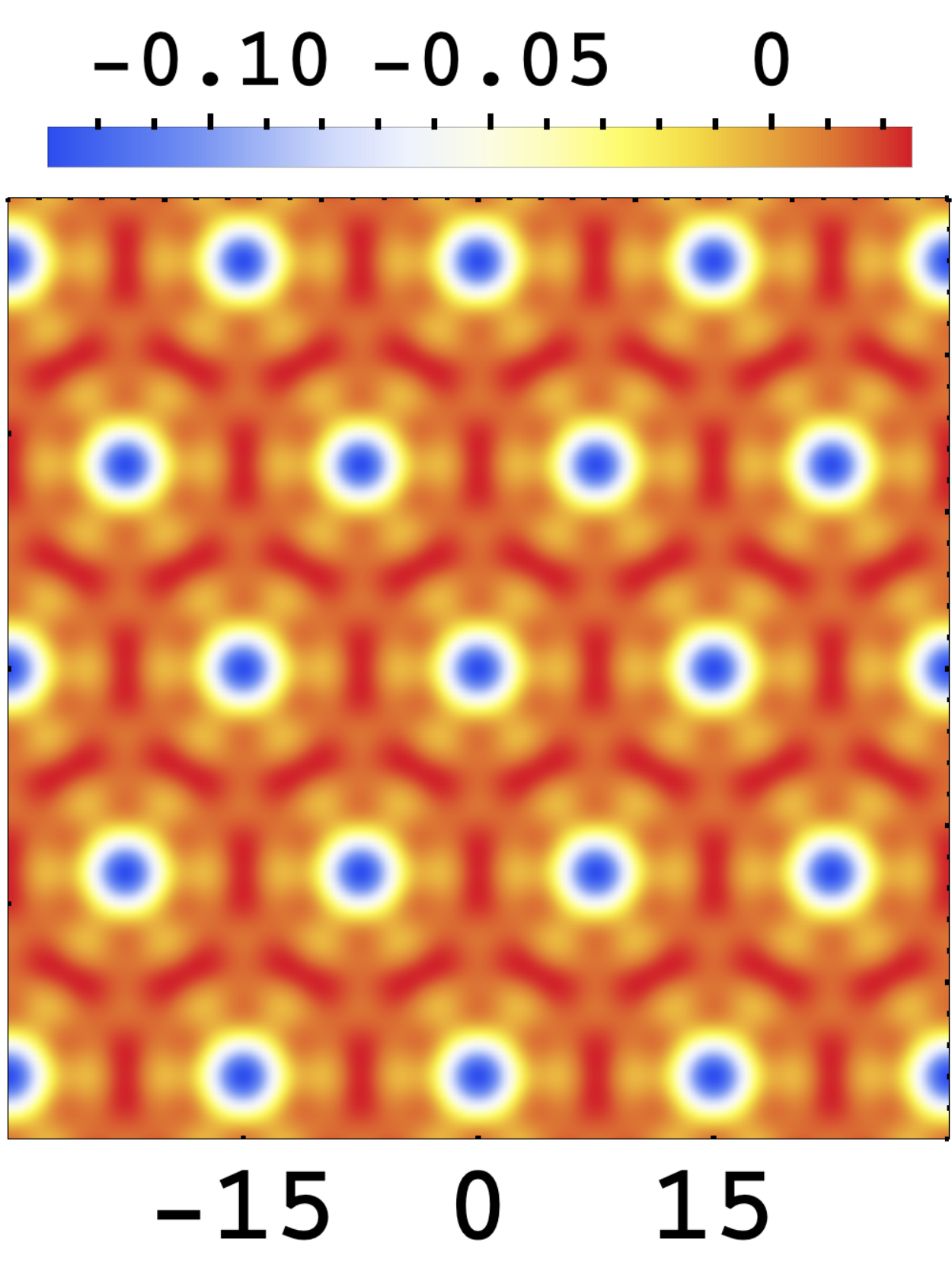}}
\hspace{-2.mm}
\resizebox{0.164\textwidth}{!}
{\includegraphics{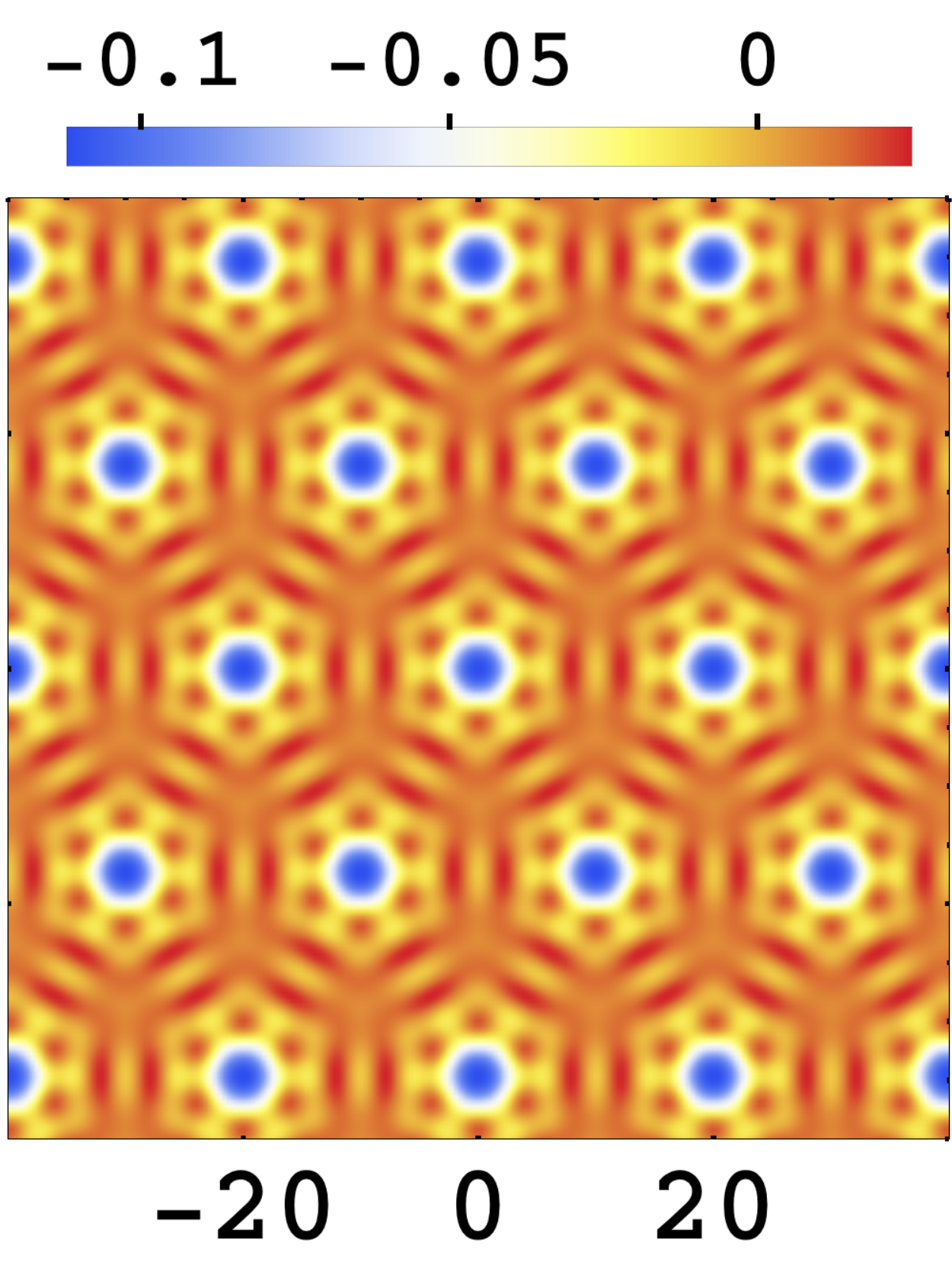}}
\hspace{-2.mm}
\resizebox{0.164\textwidth}{!}
{\includegraphics{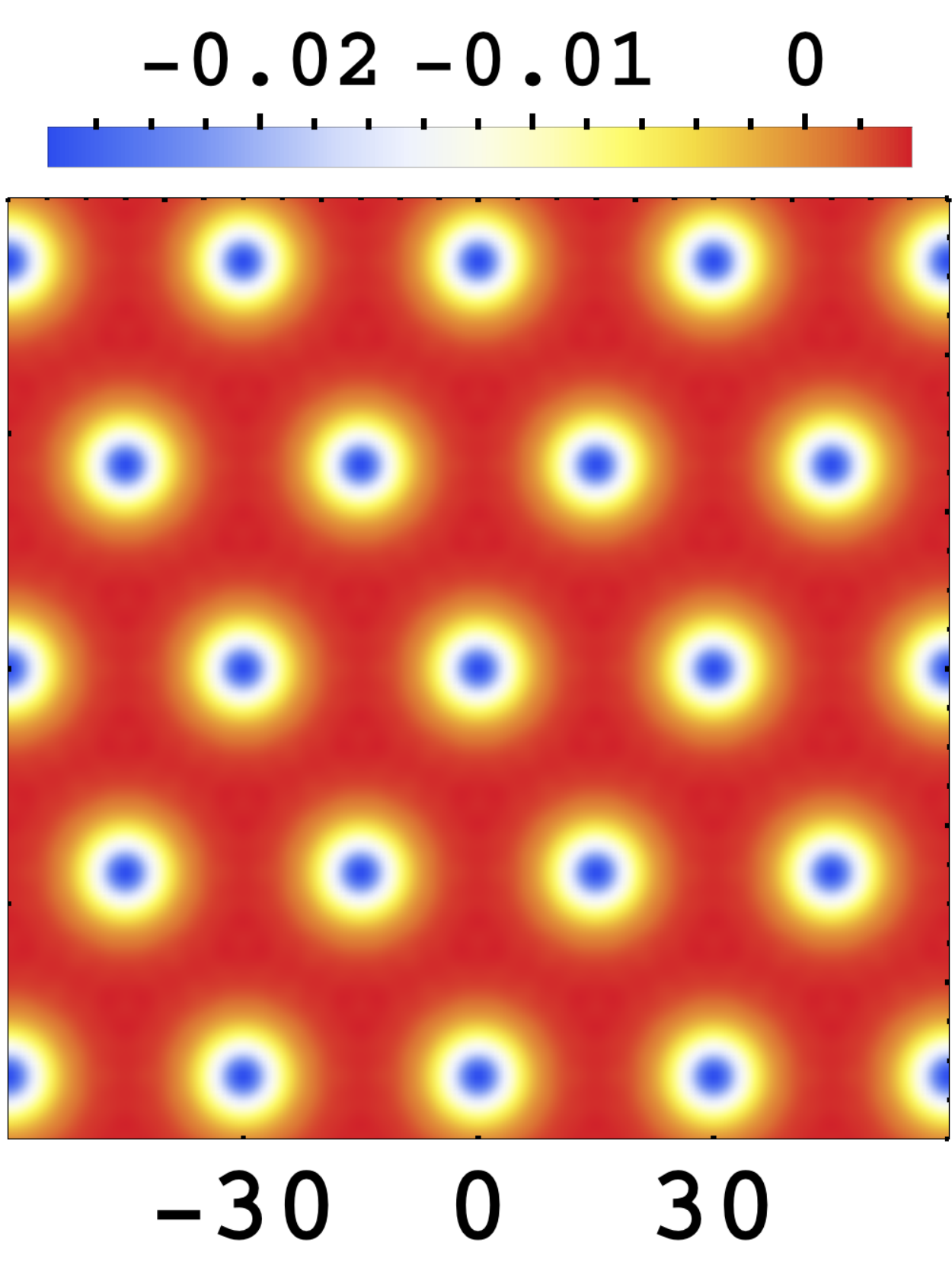}}
\vspace{-10.3mm}
\end{center}
\caption{Change in the density of the CF Fermi sea due to the presence of a WC of lattice constant $c$ in a nearby layer at a distance $d$. The three panels, from left to right, have $(d, c)=$ (3, 15), (3, 20), (6, 30). The color represents $\Delta \nu(\vec{r})$ according to scale shown on top. All lengths are in units of $l$.}
\label{WC}		
\end{figure}

As a second application of our DFT method, we consider the geometry investigated in the recent experiment of Deng {\em et al.}\cite{Deng16}, where they study commensurability oscillations of composite fermions near filling factor $\nu=1/2$ in the presence of a periodic potential produced by a Wigner crystal in a nearby layer at a distance $d$. Analogous commensurability oscillations have been observed in an antidot superlattice \cite{Kang93} and also in the presence of a one dimensional periodic potential \cite{Kamburov12,Kamburov13,Kamburov14b,Mueed15,Mueed15b}. 
We ask here how the presence of a nearby WC affects the density of composite fermions in the vicinity of $\nu=1/2$, where composite fermions form a compressible CF Fermi sea \cite{Halperin93}.  The above method is not convenient in this regime, as we have a very large number of occupied $\Lambda$Ls. We therefore work directly with Eq.~\ref{Vignale1}, setting $T=0$. We further neglect the physics of incompressibility, which should be valid for $\nu\approx1/2$, and approximate the exchange-correlation energy as $E_{\rm xc}=a \nu+b$ (from Eq.~\ref{Ajit}). Minimization with respect to the electron density gives 
\be
2a\nu(\vec r)+b+{1\over 2\pi}\int d^2 \vec{r}' {\nu(\vec r')\over\left|\vec r - \vec r'\right|}+V(\vec{r})=\mu
\ee
where we measure energies in units of $e^2/\epsilon l$ and length in units of $l$. The potential due to the WC is modeled through Eq.~\ref{Vr} with $\rho_b=-\sum_{\vec{R}} (2\pi)^{-1}e^{-|\vec{r}-\vec{R}|^2/2}$, corresponding to a Gaussian electron at each site $\vec{R}$ of a triangular lattice with lattice constant $c$. Fourier transformation gives the deviation of filling factor from its uniform value as:
\be
\Delta\nu(\vec{r})=-{4\pi\over \sqrt{3} c^2} \sum_{\vec{K}\neq 0}\frac{e^{-|\vec{K}|d - {|\vec{K}|^2\over 2} +i \vec{K}\cdot\vec{r}}}{1+2a |\vec{K}|}
\label{Dnu}
\ee
where the reciprocal lattice vectors are given by $\vec{K}=s_1 \vec{K}_1+s_2 \vec{K}_2$ with $\vec{K}_1=(2\pi /c, -2\pi / \sqrt{3} c)$, $\vec{K}_2=(0, 4\pi / \sqrt{3} c)$, and $s_1$, $s_2$ are integers. $\Delta\nu(\vec{r})$ is independent of the unperturbed filling factor (provided we are in the compressible region near $\nu=1/2$). The parameters $b$, $\mu$ and the potential due to the uniform neutralizing background only couple to $\vec{K}=0$ and thus play no role in $\Delta \nu(\vec{r})$.  Figs.~\ref{WC} and S4 (in Supplemental Material\cite{DFT-SM}) show $\Delta\nu(\vec{r})$ for several values of $d$ and $c$. An injected composite fermion sees the sum of the external and the Hartree potentials $V_{\rm H}(\vec{r})+V(\vec{r})=\mu-2a\nu(\vec{r})-b$, and is thus attracted to high density regions. Complex patterns can appear, often dominated by values of $|\vec{K}|\approx-1/2a\approx 3/2$ where the denominator becomes small, as seen in the left two panels of Fig.~\ref{WC}. In such situations, the potential experienced by an injected composite fermion is complicated and may not produce clearly identifiable geometric resonances. However, for large $d$ and large $c$ these additional patterns are suppressed by the numerator and the density $\Delta \nu(\vec{r})$ closely reflects a hexagonal lattice as seen in the right panel of Fig.~\ref{WC}, thus allowing standard commensurability oscillations. This is consistent with the experiments of Deng {\em et al.}~\cite{Deng16}, where they observe commensurability oscillations for relatively large values of $d$ and $c$ ($d\approx 6.5$ and $20 < c <60$).

We have assumed in our calculations a fully spin polarized system, as appropriate for sufficiently high magnetic fields.  It would be interesting to extend our approach to include spin and explore the possibility of spin textures at the edge\cite{Karlhede96,Lubin97,Heinonen99,Zhang13}.  Zhang, Hu and Yang have investigated precisely the model studied above by careful exact diagonalization studies\cite{Zhang13} and concluded that edge reconstruction of the 1/3 state does not involve spin reversal unless the magnetic field is very small ($<$1.0 T for GaAs). This is not surprising because, as stressed by Karlhede {\em et al.}\cite{Karlhede96}, the energetics of spin textures at the edge is closely related to that of skyrmions\cite{Sondhi93}, and calculations have shown that skyrmions at 1/3 become viable only at very low Zeeman energies\cite{Kamilla96}. We note that the Chern-Simons mean field theory of composite fermions~\cite{Halperin93,Lopez91} has also been used to treat the effect of an external periodic potential on the state in the vicinity of half filling~\cite{Zhang14c,Zhang15}; see SM for a comparison with our approach.

In summary, we have presented a new formulation of the density functional theory of the FQHE that offers a natural way of producing fractionally occupied Kohn-Sham orbitals of electrons. 
We have introduced an exchange correlation energy that is consistent with microscopic calculations, and an entropy that incorporates the physics of strong correlations. We have applied our DFT to study the physics of the FQH edge as well as to the CF Fermi sea exposed to a periodic potential.

We acknowledge financial support from the US Department of Energy under Grant No. DE-SC0005042. We are grateful to Ajit Balram, Paul Lammert, Mansour Shayegan and Giovanni Vignale for illuminating discussions.

%

\pagebreak
\appendix
\widetext 
\section{Supplemental Material}

\setcounter{figure}{0}
\setcounter{equation}{0}
\renewcommand\thefigure{S\arabic{figure}}
\renewcommand\thetable{S\arabic{table}}
\renewcommand\theequation{S\arabic{equation}}

We first provide details of our numerical methods, and discuss possible sources of numerical instability and how our approach deals with them. We then present additional results for edge reconstruction alluded to in the main article, as well as for the density profile of the 1/2 Fermi sea in the presence of a nearby Wigner crystal for a larger range of parameters. Finally, we calculate the density variation for a model previously treated by Chern-Simons mean field theory and compare the results from the two methods.

\subsubsection{\bf Numerical methods}

We numerically solve for the $\nu^*_j(\vec{r})$ by discretizing the problem. Assuming that the electron density has a rotational invariance in disk geometry, it is sufficient to discretize along the radial direction, and we take 20000 discretized points for a system with radius of $200 l$. (It should be noted that while the system has rotational symmetry, the problem remains inherently two-dimensional; e.g. the computation of the Hartree potential requires a two-dimensional integral.) We begin with an initial choice of the electron density that tracks the neutralizing background charge. (We have checked that our final results are not sensitive to this choice. We obtain the same self-consistent solution if use the density profile obtained by annealing the system at a high T.) This corresponds to a specific choice for the local CF filling $\nu^*(\vec{r})$, which, in turn, gives the local occupation of different $\Lambda$ levels, $\nu^*_j(\vec{r})$, assuming lowest CF kinetic energy. Using these initial values for $\nu^*_j(\vec{r})$ on the right hand side of Eq.~13, we compute new $\nu^*_j(\vec{r})$, while fixing the chemical potential so as to ensure the correct total charge. A new choice for $\nu^*_j(\vec{r})$  is then obtained by linearly mixing 20\% of output $\{\nu_i^*(r)\}$ into the input. (A mixing of 10\% of the output also gives the same final results.) The process is iterated until self-consistency is achieved, defined so  that the change in $\nu^*_j(\vec{r})$ is less than 0.1\% at each discrete site. 

To ensure smoothness on the scale of $l^*$, which is expected on physical grounds, we average the local CF filling factor of each Lambda level ($\nu^*_j(\vec{r})$) over a length $l_{\rm ave}$ at each step of our self-consistency loop. In our calculations shown in this article, we use $l_{\rm ave}=l^*$ (which depends on the local filling factor). The computational scheme described here is quite efficient, and the convergence is usually reached in fewer than 400 iterations, taking about an hour on a single core desktop. 

\subsubsection{\bf Sources of numerical instability}

A remark regarding numerical instabilities is in order. In previous works\cite{Ferconi95,Heinonen95}, there were two sources of instability. (i) Because the problem was formulated in terms of electrons, the filling factor $\nu$ was given by \cite{Ferconi95} (neglecting occupation of higher Landau levels, which is negligible at low temperatures)
\be 
\nu(\vec{r})=2\pi l^2 \rho(\vec{r})={1\over  e^{(V+V_{\rm H}+V_{\rm xc}-\mu)/k_BT}+1  }
\label{FGV}
\ee
which exponentially approaches either 0 or 1 in the low-$T$ limit. (ii) Second, they used the $T=0$ form of $V_{\rm xc}$. The discontinuities in $V_{\rm xc}$ resulted in numerical instabilities, because at $\nu=n/(2n\pm 1)$ an arbitrarily small increase in the density leads to a finite change in $V_{\rm xc}$. The latter source was eliminated in Ref.~\cite{Heinonen95} by replacing the discontinuities in $V_{\rm xc}$ by lines of large slopes (i.e., by introducing a small compressibility).  However, the difficulty mentioned in (i) precluded convergence in Ref.~\cite{Ferconi95} for $k_B T< 0.7 E^*_{\rm F}$ (where $E^*_{\rm F}\approx 0.1 e^2/\epsilon l$ is the CF Fermi energy).

The severity of these problems is significantly reduced in our formulation. First, $\nu_j^*$ for each $j$ approaches either 0 or 1 as $T\rightarrow 0$, which produces an integer value $\nu^*=n$ for the CF filling or a fractional value $\nu=n/(2n\pm 1)$ for the electron filling. For these filling factors, we do not need to resort to either thermal averaging or ensemble averaging. Furthermore, this reduces the fluctuations in density significantly, as the electron density fluctuates between two nearby fractions rather than between the integers 0 and 1. Second, even though we are not using it explicitly in our calculations, $V_{\rm xc}$ is T dependent, and its discontinuities are washed out by thermal smearing at $T\neq 0$ (Fig.~1 of main text). We find that with the local averaging of $\nu^*(\vec{r})$, we are able to go to temperatures as low as $10^{-2} E^*_{\rm F}$ without encountering numerical instabilities.

\subsubsection{\bf Additional results}

We next provide additional results on edge reconstruction alluded to in the main article, as well as the density profile of the 1/2 Fermi sea in the presence of a nearby Wigner crystal for a larger range of parameters.

\begin{figure}[hbt]
\hspace{-2.4mm}
\resizebox{0.2\textwidth}{!}
{\includegraphics{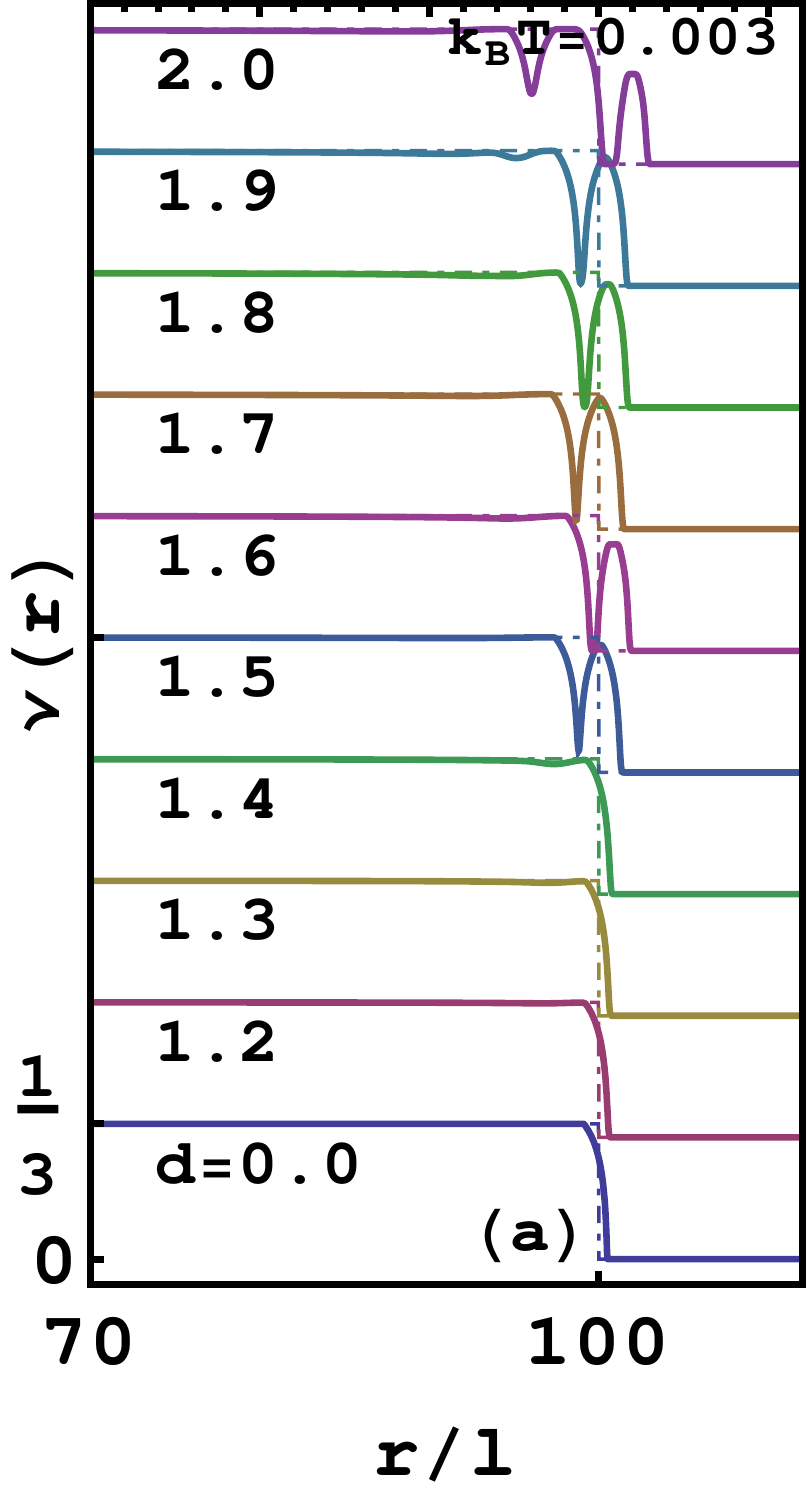}}
\resizebox{0.2\textwidth}{!}
{\includegraphics{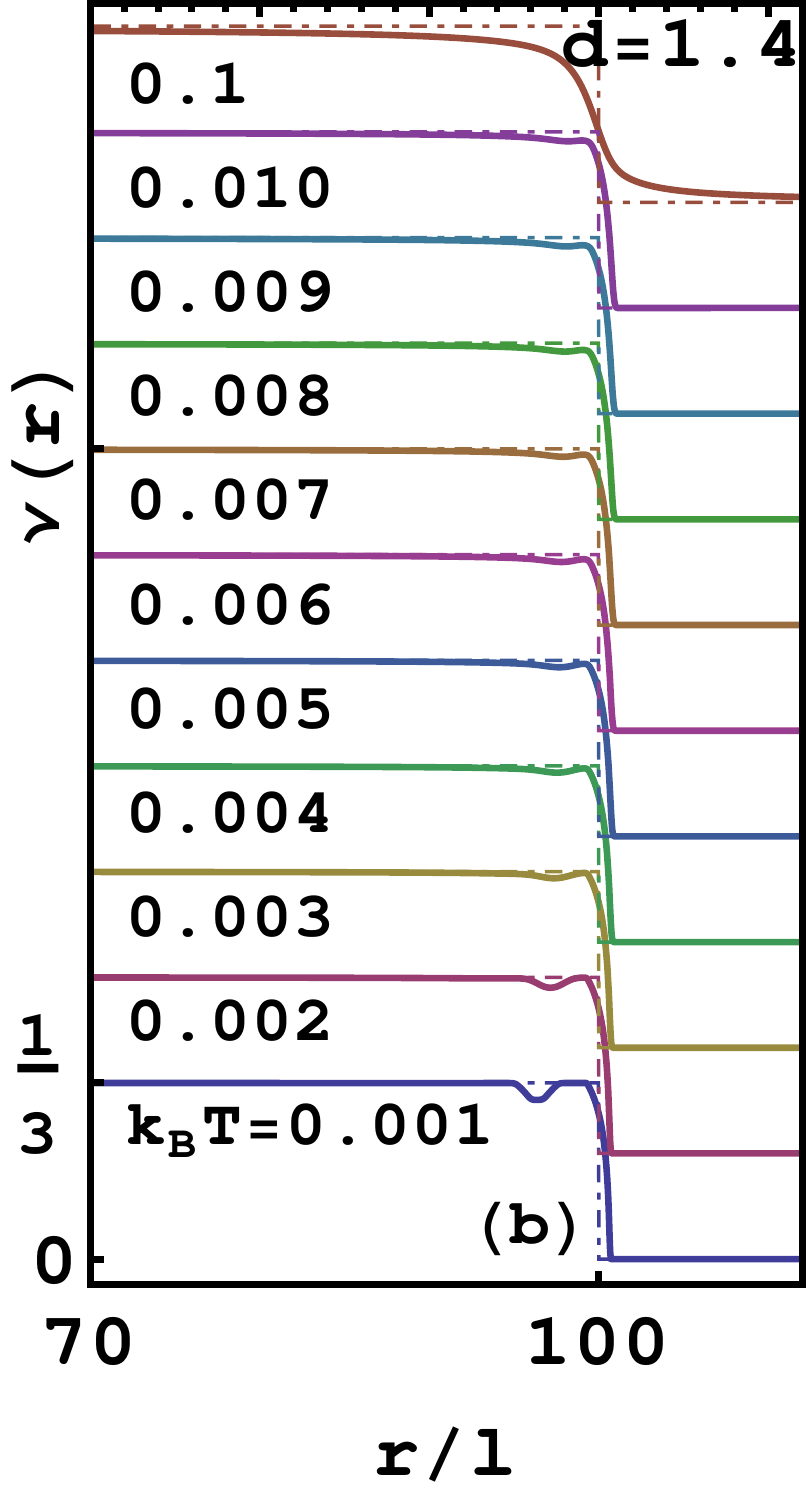}}
\resizebox{0.2\textwidth}{!}
{\includegraphics{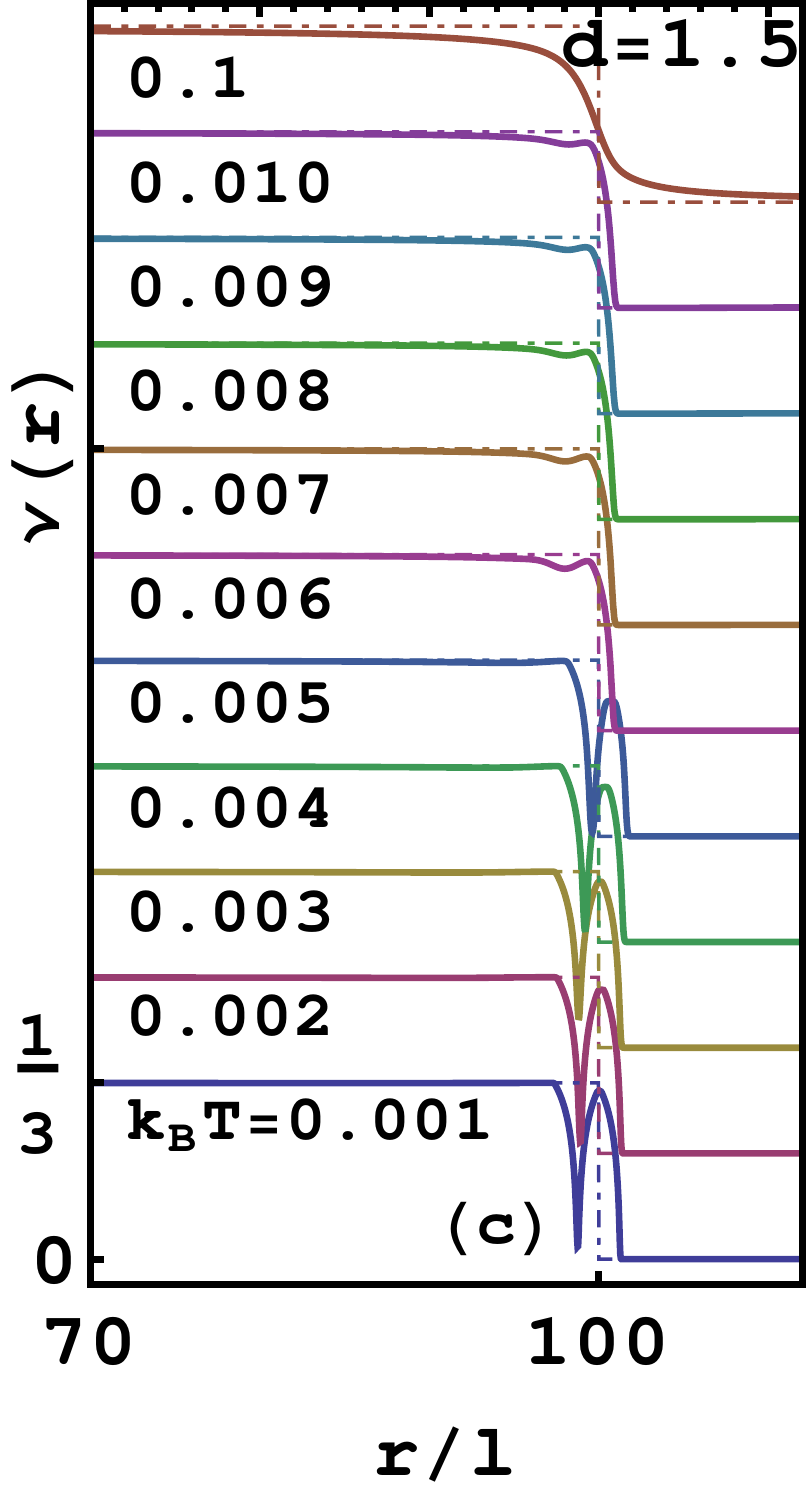}}
\resizebox{0.2\textwidth}{!}
{\includegraphics{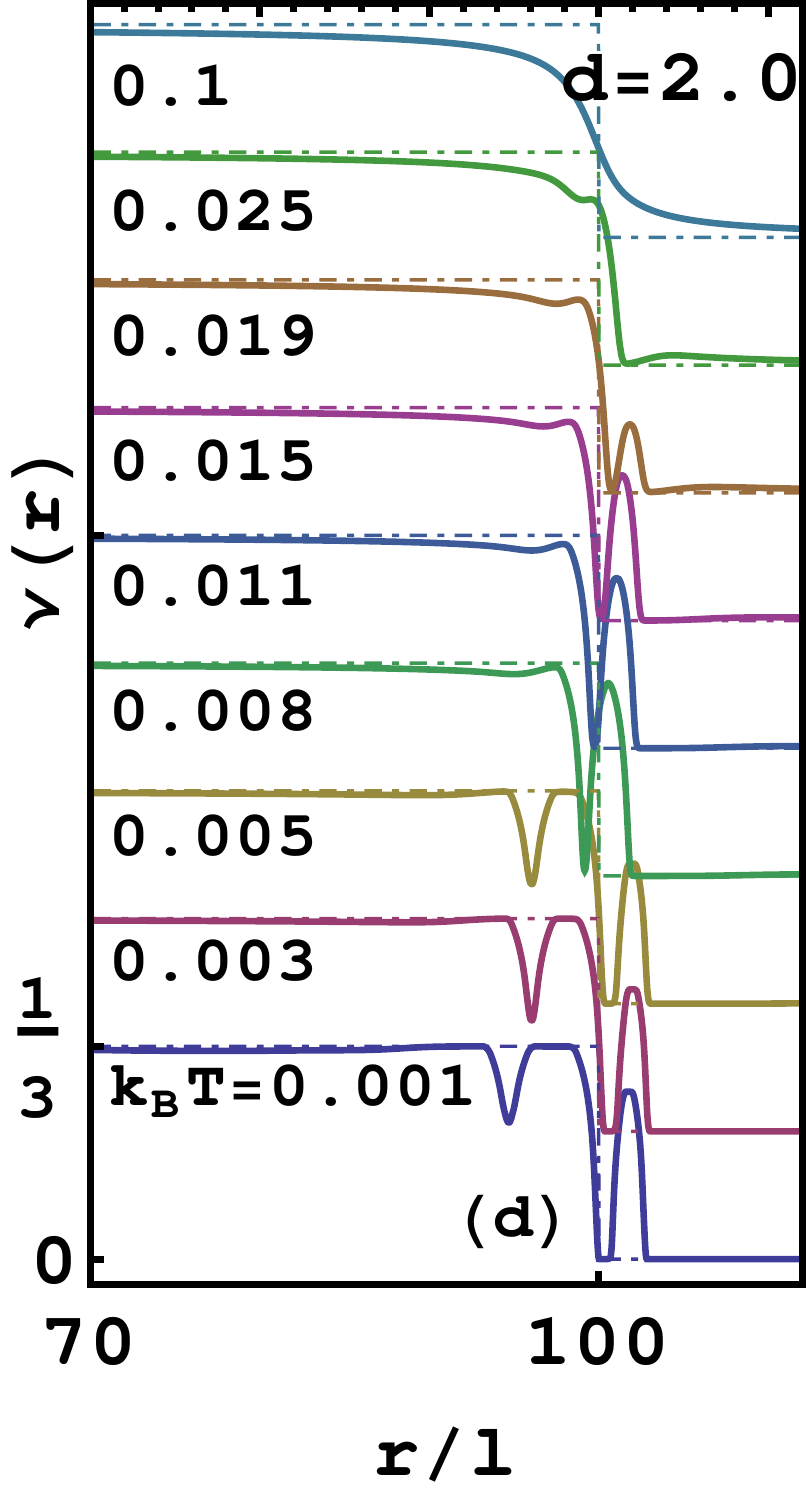}}
\\
\hspace{-4.4mm}
\resizebox{0.33\textwidth}{!}
{\includegraphics{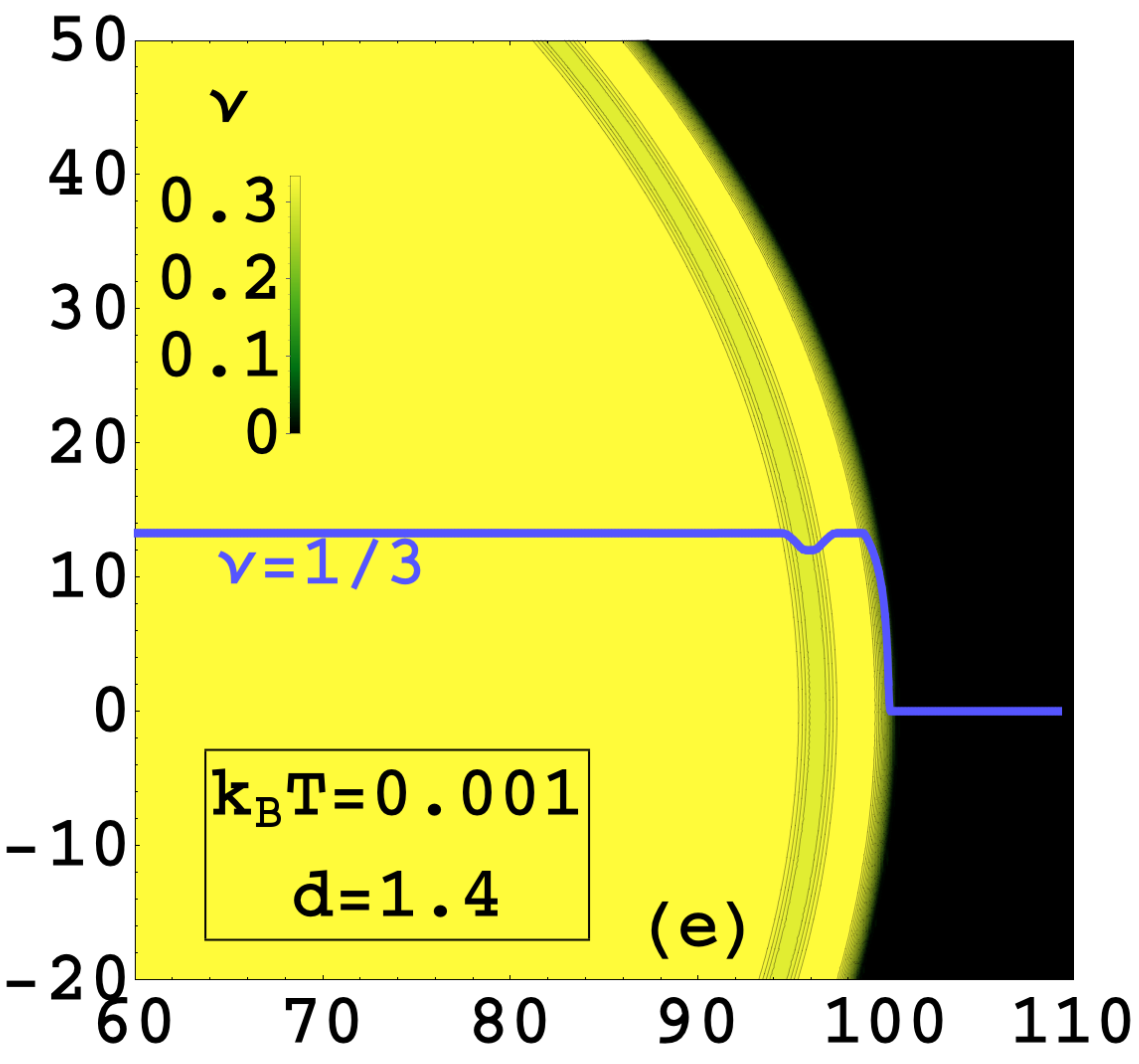}}
\hspace{-2.0mm}
\resizebox{0.33\textwidth}{!}
{\includegraphics{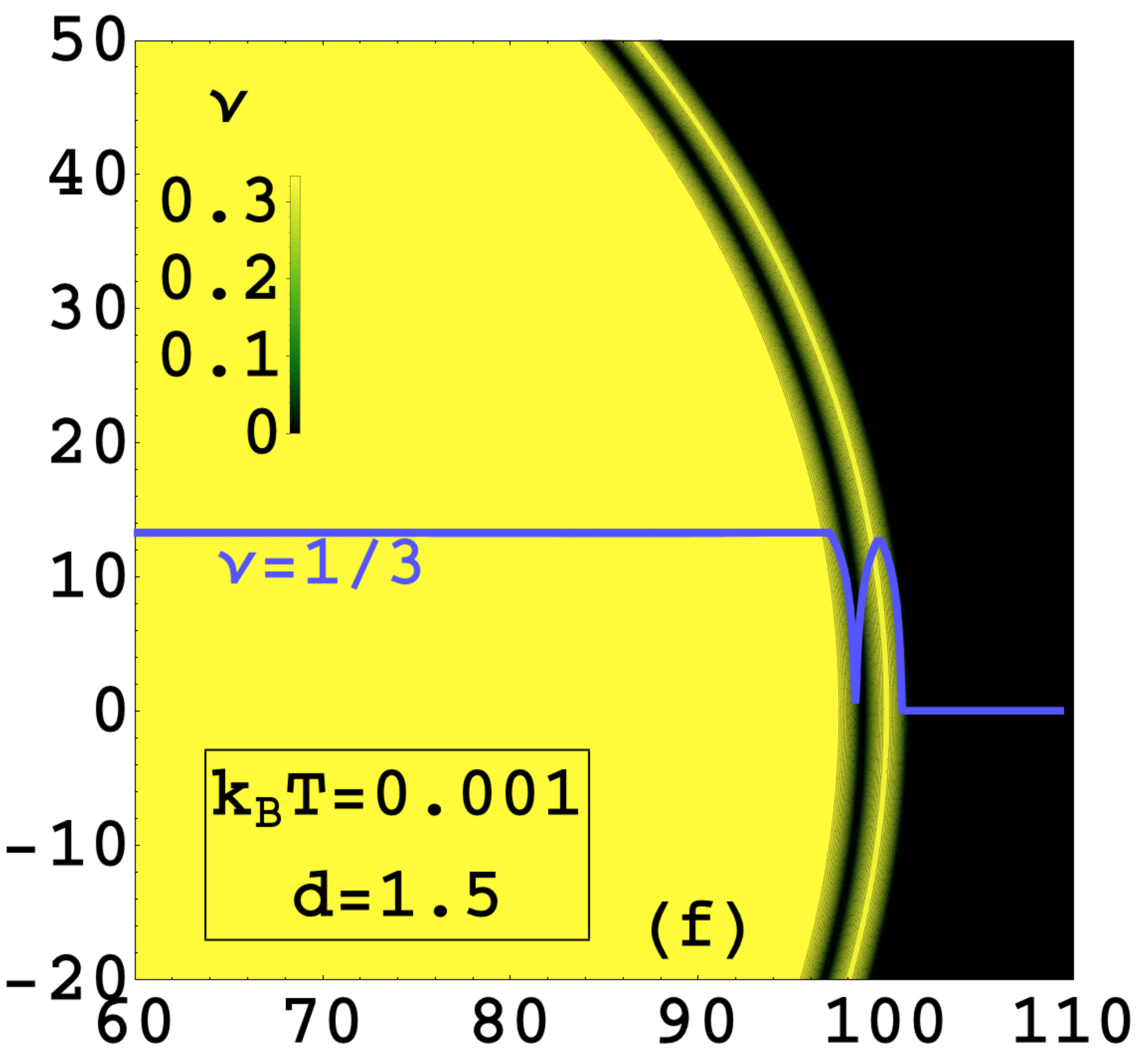}}
\hspace{-2.4mm}
\vspace{-3.3mm}
\caption{(a) Evolution of the 1/3 edge as a function of the set-back distance $d$ at a small temperature $k_{\rm B}T=0.003$. The $\nu(\vec{r})$ for successive $d$ are vertically displaced for clarity. Edge reconstruction is seen to occur at $d\approx 1.5$. (b-d) Evolution of the 1/3 edge as a function of temperature for three values of $d$. For $d=1.5 l$ the edge structure melts at $k_{\rm B}T\approx 0.006$. (e-f) The 2D density plots of the edges in the disk geometry at $k_{\rm B}T=0.001$ and set-back distances $d=1.4$ and $1.5$.} 
\label{13edge}
\end{figure}

Fig.~\ref{13edge} shows the edges of the 1/3 fractional quantum Hall states as a function of the set-back distance $d$ and temperature T. 
Fig.~\ref{13edge}(a) displays the evolution of the 1/3 edge at a low temperature as a function of the set-back distance $d$. Edge reconstruction is seen to occur at $d=1.5l$. This is consistent with exact diagonalization studies [11-14] that find an instability of the 1/3 edge at approximately the same value of the setback distance.  We take this to be a strong evidence for the validity of our approach. Figs.\ref{13edge}(b) and (c) illustrate the evolution of the edge as a function of temperature. At $d=1.4$ there is no sign of reconstructed edge down to $k_{\rm B}T=0.001$, while at $d=1.5$ edge reconstruction is washed out by $k_{\rm B}T=0.006$. Figs.\ref{13edge}(d) and (e) display two-dimensional density plots in the disk geometry to underscore how the behavior of edge changes dramatically with a small change in $d$. For completeness, we show in Fig.~S2 the edge of 3/7 FQH state as a function of temperature for a set-back distance slightly below the critical value.

Fig.~S3 shows the self-consistent $\Lambda$L energies $\epsilon_j^*(\vec{r})$ for $\nu=3/7$ with setback distance $d=2.2$. Three different values of temperature are chosen for illustration. 

Fig.~S4 displays the density profile $\Delta\nu$ of the composite fermion Fermi sea under the influence of a nearby Wigner crystal. A wider range of the setback distance $d$ and the lattice constant $c$ are chosen. The rich structures seen at low $d$ disappear for $d\gtrsim$5.

\subsubsection{\bf Comparison with a Chern-Simons study}

Y.H. Zhang and J.R. Shi\cite{Zhang14c} have studied the CF Fermi liquid near filling factor $1\over2$ superimposed by a weak hexagonal period potential employing an effective Chern-Simons theory. They find that composite fermions experience a staggered effective magnetic field $B^*$ and the system becomes a quantum anomalous Hall insulator for CFs when each unit cell of the external potential contains an integer number of electrons. Here we study this model using the formalism presented in the main paper, namely Eq. 19, which is appropriate for dealing with the CF Fermi sea near $\nu={1\over2}$.

The periodic potential considered in Ref. ~\cite{Zhang14c} is given by (converting to our convention for the units and symbols):
\be
U(\vec r)=-{V_0\over2}\left\{\rm{cos}\left((\vec b_1+\vec b_2).\vec r-{\pi\over3}\right)+\rm{cos}\left(\vec b_1.\vec r+{\pi\over3}\right)+\rm{cos}\left(\vec b_2.\vec r+{\pi\over3}\right)\right\}+{3V_0\over2}
\label{Ur}
\ee 
where $\vec b_1=4\pi/(\sqrt{3}c)(0,1)$ and $\vec b_2=(2\pi/\sqrt{3}c)(\sqrt{3},-1)$, $c=3.81$ and $V_0=0.108$. Solving Eq. 19 (main paper) by Fourier transform yields the following filling factor deviation from its uniform value:
\be
\Delta\nu(\vec r)={-1\over2a+{1\over|\vec b_1|}}(U(\vec r)-{3V_0\over2})
\ee
Fig.~\ref{ChernS} shows the two-dimensional density plot of $\Delta\nu$. We note that our result is modified by the exchange correlation energy of composite fermions (through the parameter $a$); in contrast, Ref.~\cite{Zhang14c} does not include the exchange correlation energy. Also, the large change in $\Delta \nu$ (note $|\Delta \nu|>0.5$ is not physically acceptable) suggests that the potential is not weak. The effective magnetic field is given by $B^*=B(1-2\nu)=-2B\Delta\nu=-{\phi_0\over\pi l^2}\Delta\nu$. In Ref.~\cite{Zhang14c} the magnetic field is quoted as the number of flux quanta through area $c^2\over3$, which corresponds to $c^2B^*/3\phi_0=-c^2\Delta\nu /3\pi l^2\approx -1.54 \Delta \nu$. In our calculation, $\Delta \nu$ changes from $-0.5$ to 1.4, which translates into a significantly larger variation in $B^*$ than that found in Ref.~\cite{Zhang14c}.

 \begin{figure}[hbt]
\resizebox{0.2\textwidth}{!}
{\includegraphics{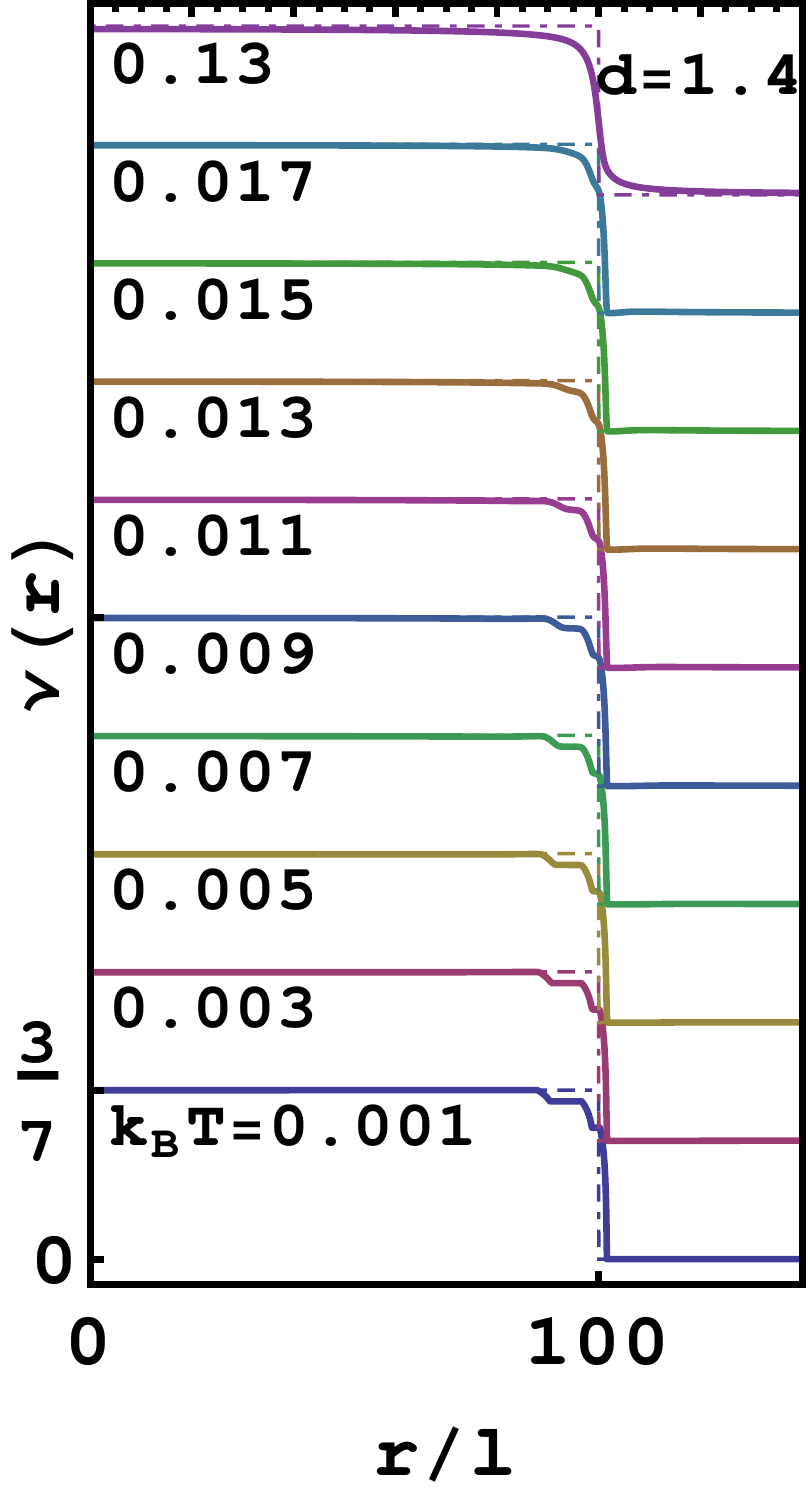}}
\hspace{-2.4mm}

\caption{The evolution of 3/7 edge as a function of temperature for the set-back distance $d=1.4$, which is between the two values provided in Fig.~2 (b) and (c) of the main paper. It shows incompressible stripes at $\nu=2/5$ and $1/3$ but no reconstructed edge.}
\end{figure}

 \begin{figure}[hbt]
{\includegraphics[scale=0.4]{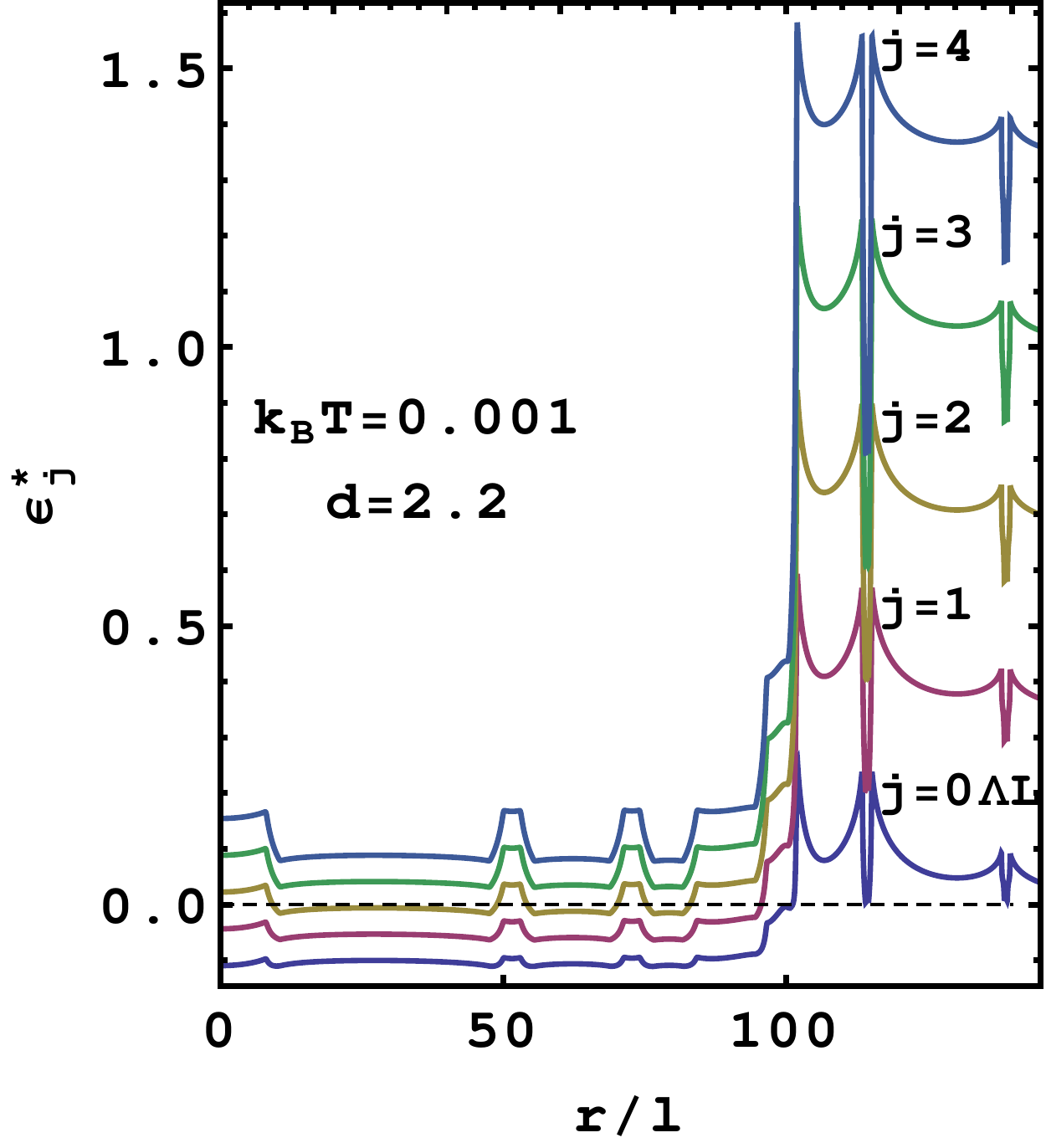}}
{\includegraphics[scale=0.4]{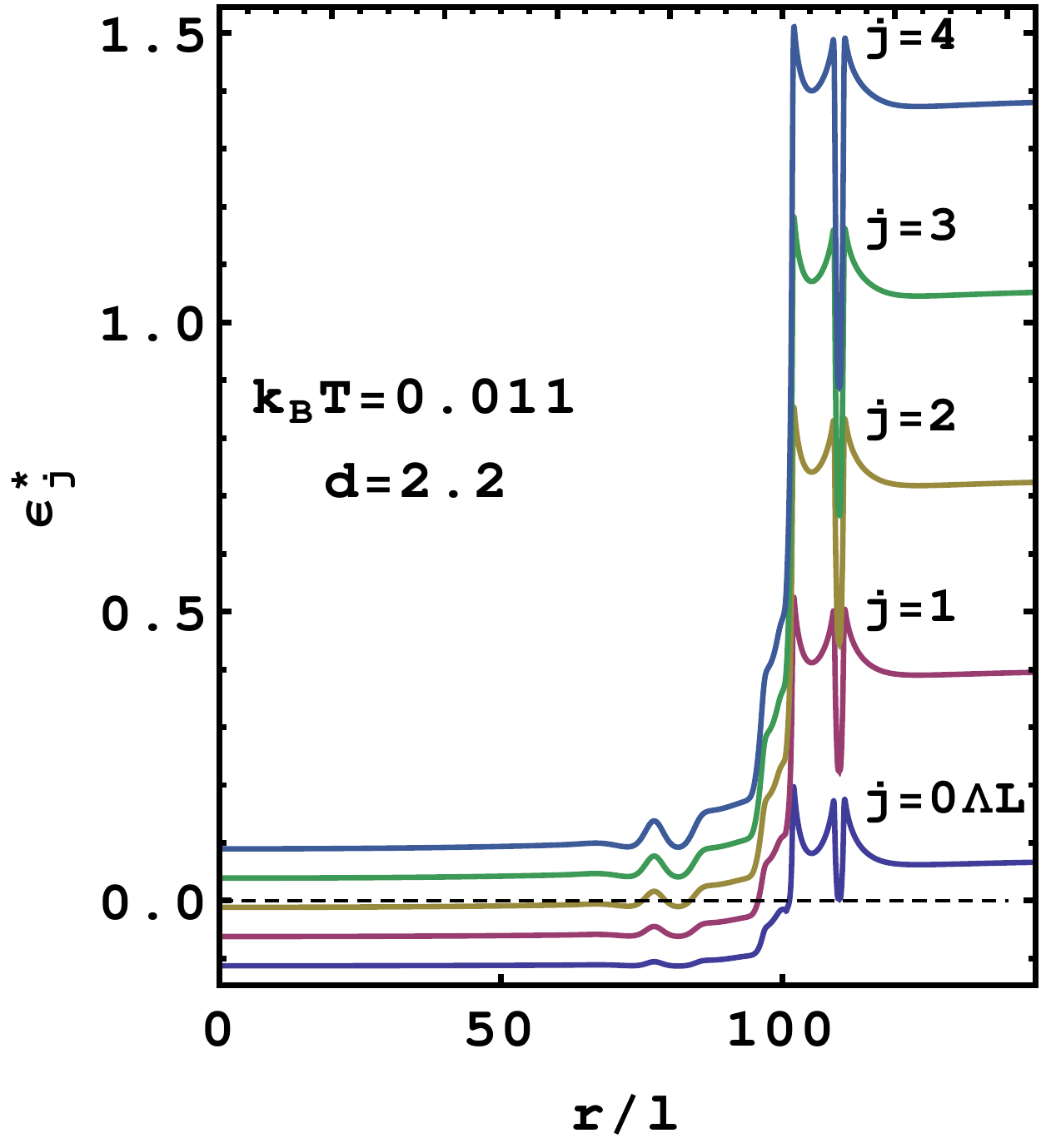}}
{\includegraphics[scale=0.4]{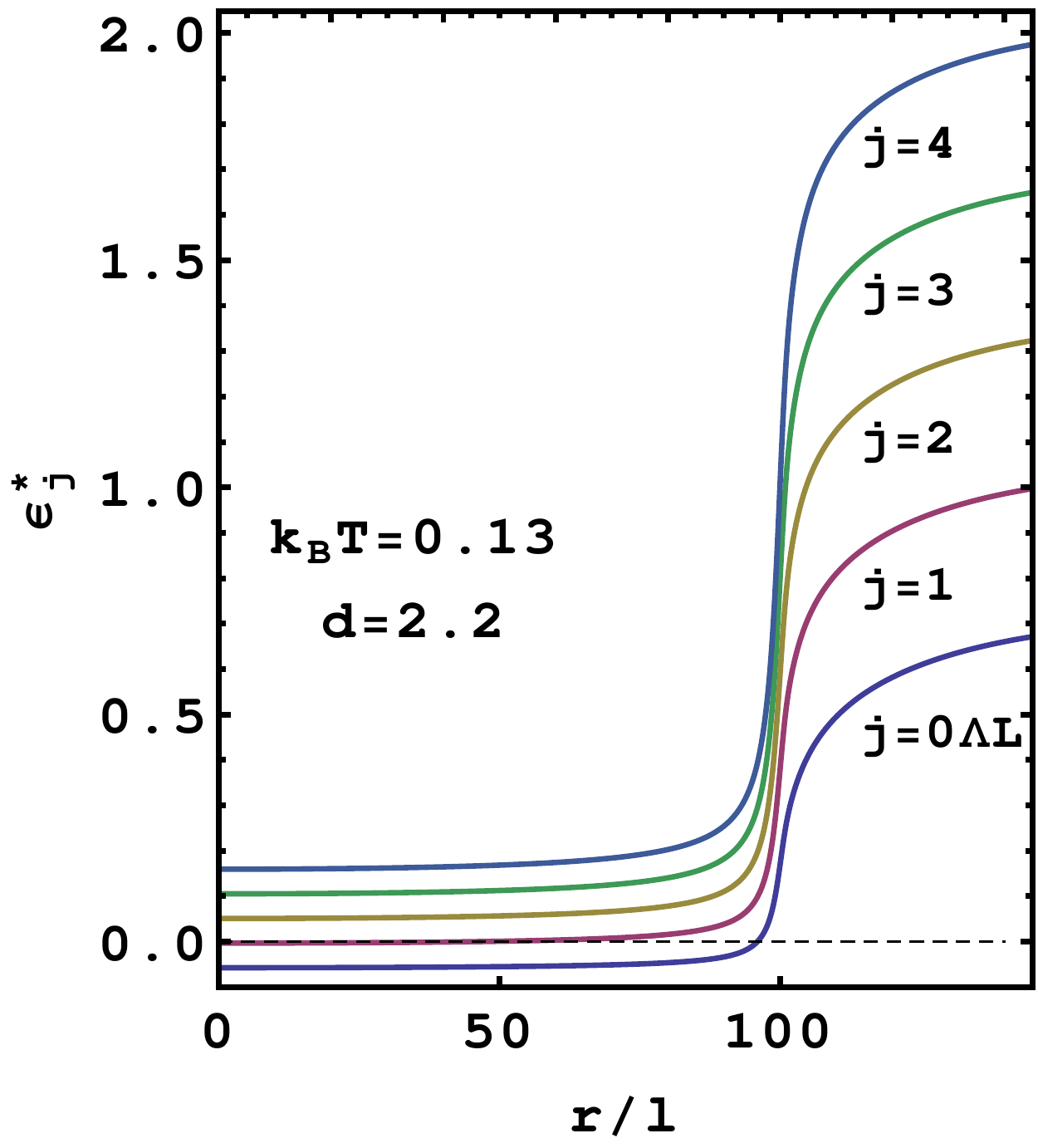}}
\caption{This figure shows the self-consistent $\Lambda$L energies $\epsilon^*_j(\vec{r})$ for the lowest 5 $\Lambda$Ls. The three panels are for temperatures $k_{\rm B}T=$ 0.001, 0.011 and 0.13. All panels are for $\nu=3/7$ and $d=2.2$. The energies are given in units of $e^2/\epsilon l$. The $\Lambda$L filling factors $\nu_j^*(\vec{r})$ are given by Eq.~14; aside from thermal smearing, the $\Lambda$Ls below zero energy (marked by the horizontal dashed line) are occupied and those above unoccupied.}
\end{figure}

\begin{figure*}[hbt]
\begin{center}			
\resizebox{0.7\textwidth}{!}
{\includegraphics{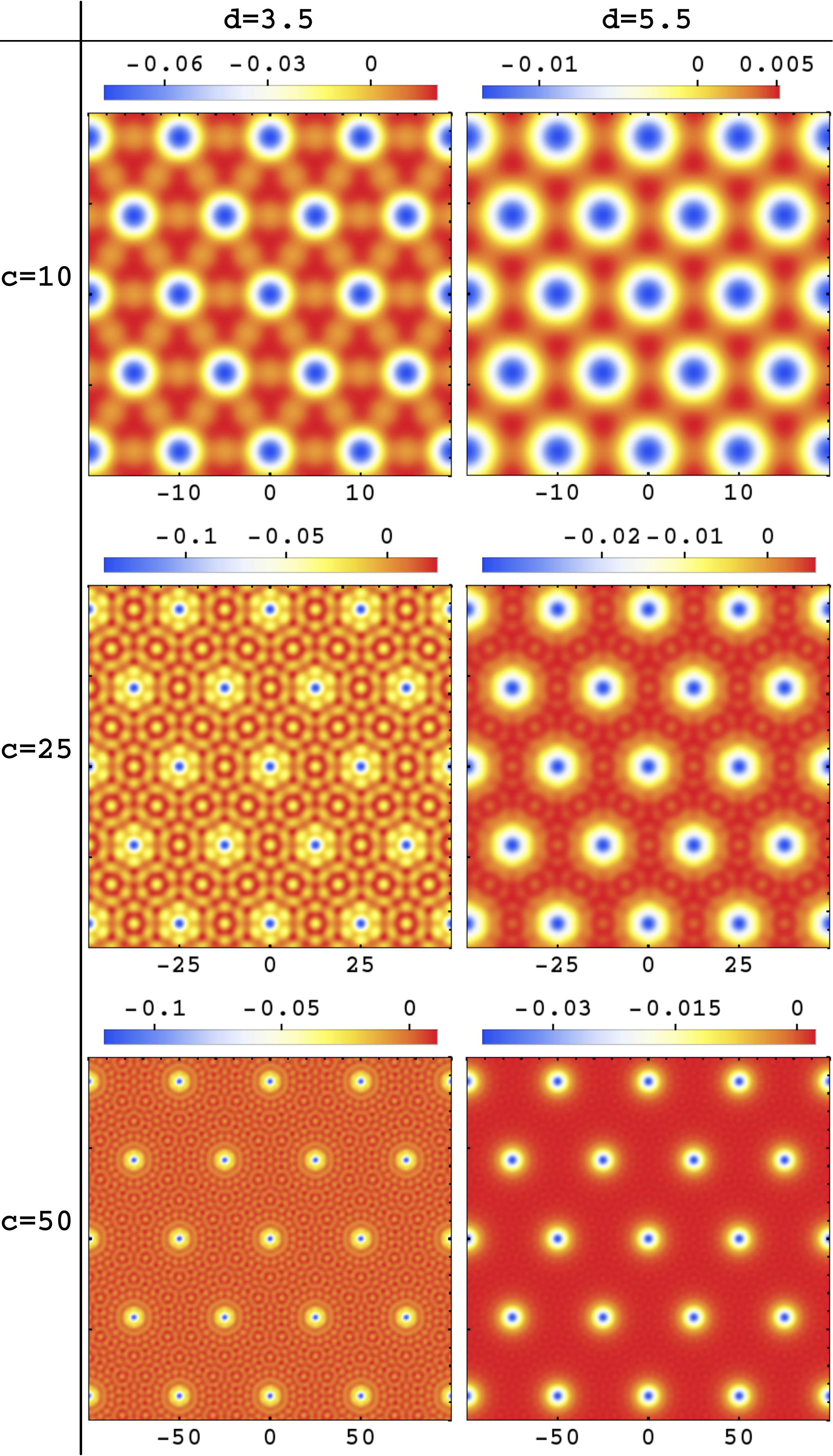}}
\end{center}
\caption{Change in the density of the CF Fermi sea due to interaction with a Wigner crystal of lattice constant $c$ in a two-dimensional layer at a distance $d$. The color represents the deviation of filling factor $\Delta \nu(\vec{r})$ from its uniform value according to the scale shown on top. The rich structures shown at low $d$ disappear when $d\gtrsim$5.
}
\label{Vxc_plot}		
\end{figure*}	

 \begin{figure}[hbt]
\vspace{-2.4mm}
\resizebox{0.4\textwidth}{!}
{\includegraphics{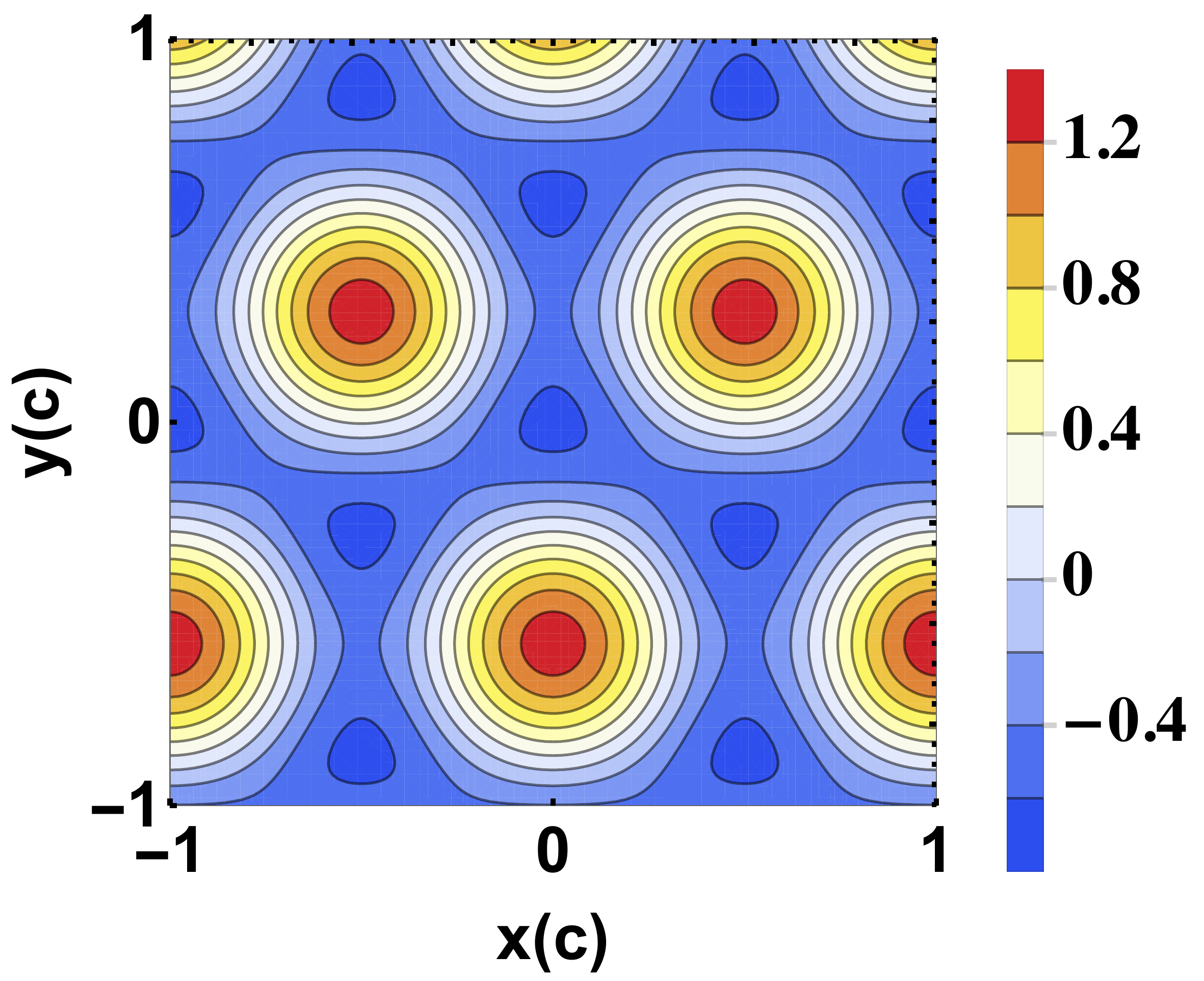}}
\vspace{-3.4mm}
\caption{Change in the filling factor $\Delta\nu(\vec r)$ due to the presence of a  hexagonal potential $U(\vec r)$ of Eq.~\ref{Ur}. The lengths $x$ and $y$ are quoted in units of the lattice constant $c=3.81 l$.}
\label{ChernS}
\end{figure}

\end{document}